\documentclass[iop]{emulateapj}
\usepackage{rotating}
\pdfoutput=1
\usepackage{graphicx}
\usepackage{epstopdf}
\slugcomment{Draft: \today}

\shorttitle{Dipole and Quadrupole Mixing}
\shortauthors{A. Finley \& S. Matt}

\begin{document}


\title{The Effect of Combined Magnetic Geometries on Thermally Driven Winds I: \\
 Interaction of Dipolar and Quadrupolar Fields}


\author{Adam J. Finley* \& Sean P. Matt}
\affil{University of Exeter (UK), Department of Physics \& Astronomy,
              Devon, Exeter, EX4 4QL}
\email{*af472@exeter.ac.uk}



\begin{abstract}
Cool stars with outer convective envelopes are observed to have magnetic fields with a variety of geometries, which on large scales are dominated by a combination of the lowest order fields such as the dipole, quadrupole and octupole modes. Magnetised stellar wind outflows are primarily responsible for the loss of angular momentum from these objects during the main sequence. Previous works have shown the reduced effectiveness of the stellar wind braking mechanism with increasingly complex, but singular, magnetic field geometries. In this paper, we quantify the impact of mixed dipolar and quadrupolar fields on the spin-down torque using 50 MHD simulations with mixed field, along with 10 of each pure geometries. The simulated winds include a wide range of magnetic field strength and reside in the slow-rotator regime. We find that the stellar wind braking torque from our combined geometry cases are well described by a broken power law behaviour, where the torque scaling with field strength can be predicted by the dipole component alone or the quadrupolar scaling utilising the total field strength. The simulation results can be scaled and apply to all main-sequence cool stars. For Solar parameters, the lowest order component of the field (dipole in this paper) is the most significant in determining the angular momentum loss. 
\end{abstract}






\section{Introduction}
The spin down of cool stars ($M_*\lesssim 1.3M_{\sun}$) is a complex function of mass and age, as shown by the increasing number of rotation period measurements for large stellar populations \citep{barnes2003rotational, irwin2009ages, barnes2010simple, agueros2011factory, meibom2011color, mcquillan2013measuring, bouvier2014angular, stauffer2016rotation, 2017ApJ...835...16D}. Observed properties of these stars show a wide range of mass loss rates, coronal temperatures, field strengths and geometries, which all connect with stellar rotation to control the loss of angular momentum (\citealp{reiners2012radius}; \citealp{gallet2013improved}; \citealp{van2013fast}; \citealp{brown2014metastable}; \citealp{matt2015mass}; \citealp{gallet2015improved}; \citealp{amard2016rotating}; \citealp{blackman2016minimalist}; See et al. in prep). Despite the wide range of interlinking stellar properties an overall trend of spin down with an approximately Skumanich law is observed at late ages; $\Omega_* \propto \tau^{-0.5}$ \citep{skumanich1972time, soderblom1983rotational}. 

For Sun-like stars on the main sequence, the spin-down process is governed primarily by their magnetised stellar winds which remove angular momentum over the star's lifetime. \cite{parker1958dynamics} originally posited that stellar winds must exist due to the thermodynamic pressure gradient between the high temperature corona and interplanetary space. Continued solar observations have constrained theoretical models for the solar wind to a high degree of accuracy \citep{van2014alfven, usmanov2014three, oran2015steady}. Recent models of the solar wind are beginning to accurately reproduce the energetics within the corona and explain the steady outflow of plasma into the Heliosphere (e.g. \citealp{grappin1983dependence}; \citealp{van2010data}; \citealp{pinto2016flux}). The wind driving is now known to be much more complex than a thermal pressure gradient, with authors typically heating the wind through the dissipation of Alfv\'en waves in the corona. Other cool stars are observed with x-ray emissions indicating hot stellar coronae like that of the Sun \citep{rosner1985stellar, hall2007activity, wright2004chromospheric, wolk2005stellar}. Similar stellar winds and wind heating mechanisms are therefore expected to exist across a range of Sun-like stars. Assuming equivalent mass loss mechanisms, results from the Solar wind are incorporated into more general stellar wind modelling efforts \citep[e.g.][]{cohen2014grid, alvarado2016simulating}. 

Detailed studies of wind driving physics remain computationally expensive to run, so are usually applied on a case-by-case basis. How applicabile the heating physics gained from modelling the Solar wind is to other stars still in question. With the reliability of such results even for the global properties of a given star in question, large parameter studies with simpler physics remain useful. A more general method can allow for parametrisations which are more appropriate to the variety of stellar masses and rotation periods found in observed stellar populations. Parker-type solutions remain useful for this due to their simplicity and versatility \citep{parker1965dynamical, mestel1968magnetic, sakurai1990magnetohydrodynamic, keppens1999numerical}. In these solutions, wind plasma is accelerated from the stellar surface and becomes transonic at the sonic surface. With the addition of magnetic fields the wind also become trans-alfv\'enic, i.e faster than the Alfv\'en speed, at the Alfv\'en surface.  \cite{weber1967angular} showed for a one-dimensional magnetised wind that the Alfv\'en radius represented a lever arm for the spin-down torque. Since the introduction of this result, many researchers have produced scaling laws for the Alfv\'en radius (\citealp{mestel1984angular}; \citealp{kawaler1988angular}; \citealp{matt2008accretion}; \citealp{matt2012magnetic}; \citealp{ud2009dynamical}; \citealp{pinto2011coupling}; \citealp{reville2015effect}; Pantolmos. in prep) all of which highlight the importance of the magnetic field strength and mass loss rate in correctly parametrising a power law dependence. In such formulations, the mass loss rate is incourporated as a free parameter as the physical mechanisms which determines it are not yet completely understood. Measuring the mass loss rate from Sun-like stars is particularly difficult due to the wind's tenuous nature and poor emission. \cite{wood2004astrospheres} used Lyman-$\alpha$ absorption from the interaction of stellar winds and their local interstellar medium to measure mass loss rates, but the method is model-dependent and only available for a few stars. Theoretical work from \cite{cranmer2011testing} predicts the mass loss rates from Sun-like stars, but it is uncertain if the physics used within the model scales correctly between stars. Therefore, parameter studies where the mass loss rate is an unknown parameter are needed.

In addition to the mass loss rate, the angular momentum loss rate is strongly linked with the magnetic properties of a given star. Frequently researchers assume the dipole component of the field to be the most significant in governing the global wind dynamics \citep[e.g. ][]{ustyugova2006propeller, zanni2009mhd, gallet2013improved, cohen2014grid, gallet2015improved, matt2015mass, johnstone2015stellar}. Zeeman Doppler Imaging (ZDI) studies \citep[e.g.][]{morin2008large, petit2008toroidal, fares2009magnetic, vidotto2014stellar, jeffers2014e, see2015energy, see2016connection, folsom2016evolution, hebrard2016modelling, see2016studying}, provide information on the large scale surface magnetic fields of active stars. Observations have shown stellar magnetic fields to be much more complex than simple dipoles, containing combinations of many different field modes. ZDI is a topographic technique typically decomposes the field at the stellar surface into individual spherical harmonic modes. The 3D field geometry can then be recovered with field extrapolation techniques using the ZDI map as an inner boundary. Several studies have considered how these observed fields affect the global wind properties. Typically used to determine an initial 3D field solution, then a magnetohydrodynamics code evolves this initial state in time until a steady state solution for the wind and magnetic field geometry is attained \citep[e.g.][]{vidotto2011understanding, cohen2011dynamics, garraffo2016space, reville2016age, alvarado2016simulating, nicholson2016temporal, do2016magnetic}. These works are less conducive to the production of semi-analytical formulations, as the principle drivers of the spin-down process are hidden within complex field geometries, rotation and wind heating physics. 

A few studies show systematically how previous torque formulations depend on magnetic geometry using single modes. \cite{reville2015effect} explored thermally driven stellar winds with dipolar, quadrupolar and octupolar field geometries. They concluded that higher order field modes produce a weaker torque for the same field strength and mass loss, which is supported by results from \cite{garraffo2016missing}. Despite these studies and works like them, only one study has systematic scaled the mass loss rate for a mixed field geometry field \citep{strugarek2014modelling}. However, the aforementioned studies of the angular momentum loss from Sun-like stars have yet to address the systematic addition of individual spherical harmonic field modes. 

Mixed geometry fields are observed within our closest star, the Sun, which undergoes a 11 year cycle oscillating between dipolar and quadrupolar field modes from cycle minimum to maximum respectively \citep{derosa2012solar}. Observed Sun-like stars also exhibit a range of spherical harmonic field combinations. Simple magnetic cycles are observed using ZDI, both HD 201091 \citep{saikia2016solar} and HD 78366 \citep{morgenthaler2012long} show combinations of the dipole, quadrupole and octupole field modes oscillating similarly to the solar field. Other cool stars exist with seemingly stochastic changing field combinations \citep{petit2009polarity, morgenthaler2011direct}. Observed magnetic geometries all contain combinations of different spherical harmonic modes with a continuous range of mixtures, it is unclear what impact this will have on the braking torque.

In this study we will investigate the significance of the dipole field when combined with a quadrupolar mode. We focus on these two field geometries, which are thought to contribute in anti-phase to the solar cycle and perhaps more generally to stellar cycles in cool stars. Section 2 covers the numerical setup with a small discussion of the magnetic geometries for which we develop stellar wind solutions. Section 3 presents the main simulation results, including discussion of the qualitative wind properties and field structure, along with quantitative parametrisations for the stellar wind torque. Here we also highlight the dipole's importance in the braking, and introduce an approximate scaling relation for the torque. Finally in Section 4 we focus on the magnetic field in the stellar wind, first a discussion of the overall evolution of the flux, then a discussion of the open flux and opening radius within our simulations. Conclusions and thoughts for further work can then be found in Section 5. The Appendix contains a short note on the wind acceleration profiles of our wind solutions.

\section{Simulation Method}
\subsection{Numerical Setup}
This work uses the magnetohydrodynamics (MHD) code PLUTO \citep{mignone2007pluto, mignone2009pluto}, a finite-volume code which solves Riemann problems at cell boundaries in order to calculate the flux of conserved quantities through each cell. PLUTO is modular by design, capable of interchanging solvers and physics during setup. The present work uses a diffusive numerical scheme, the solver of Harten, Lax, and van Leer, HLL \citep{einfeldt1988godunov}, which allows for greater numerical stability in the higher strength magnetic field cases. The magnetic field solenoidality condition ($\nabla\cdot{\bf B}=0$) is maintained using the Constrained Transport method (See \cite{toth2000b} for discussion). 

The MHD equations are solved in a conservative form, with each equation relating to the conservation of mass, momentum and energy, plus the induction equation for magnetic field,
\begin{eqnarray}
\frac{\partial\rho}{\partial t} + \nabla\cdot\rho{\bf v} &=& 0,\\
\frac{\partial {\bf m}}{\partial t} + \nabla\cdot({\bf mv - BB +I}p_T) &=& \rho{\bf a},\\
\frac{\partial E}{\partial t} + \nabla\cdot((E+p_T){\bf v - B(v\cdot B})) &=& {\bf m \cdot a} ,\\
\frac{\partial{\bf B}}{\partial t} + \nabla \cdot ({\bf vB - Bv}) &=& 0 .
\end{eqnarray}
Here $\rho$ is the mass density, $\bf v$ is the velocity field, $\bf a$ is the gravitational acceleration, $\bf B$ is the magnetic field\footnote{The PLUTO code operates with a factor of $1/\sqrt{4\pi}$ absorbed into the normalisation of B. Tabulated parameters are given in cgs units with this factor incorporated.}, $p_T=p+B^2/8\pi$ is the combined thermal and magnetic pressure, $\bf m$ is the momentum density given by $\rho{\bf v}$ and $E$ is the total energy density. The energy of the system is written as $E = \rho\epsilon +{\bf m}^2/(2\rho) + {\bf B}^2/2$, with $\epsilon$ representing the internal energy per unit mass of the fluid. $\bf I$ is the identity matrix. A polytropic wind is used for this study, such that the closing equation of state takes the form $\rho\epsilon = p/(\gamma -1)$ where $\gamma$ represents the polytropic index. 

We assume the wind profiles to be axisymmetric and solve the MHD equations using a spherical geometry in 2.5D, i.e. our domain contains two spatial dimensions ($r, \theta$) but allows for 3D axisymetric solutions for the fluid flow and magnetic field using three vector components ($r, \theta, \phi$). The domain extends from one stellar radius ($R_*$) out to $60R_*$ with a uniform grid spacing in $\theta$ and a geometrically stretched grid in $r$, which grows from an initial spacing of $0.01R_*$ to $1.08R_*$ at the outer boundary. The computational mesh contains $N_r\times N_{\theta}=256\times 512$ grid cells. These choices allow for the highest resolution near the star, where we set the boundary conditions that govern the wind profile in the rest of the domain. 

Initially a polytropic parker wind \citep{parker1965dynamical, keppens1999numerical} with $\gamma=1.05$ fills the domain, along with a super-imposed background field corresponding to our chosen magnetic geometry and strength. During the time-evolution, the plasma pressure, density, and poloidal components of the magnetic field ($B_r,B_\theta$) are held fixed at the stellar surface, whilst the poloidal components of the velocity ($v_r,v_\theta$) are allowed to evolve in response to the magnetic field (the boundary is held with $dv_r/dr=0$ and  $dv_{\theta}/dr=0$). We then enforce the flow at the surface to be parallel to the magnetic field ($\bf v || B$). The star rotates as a solid body, with $B_{\phi}$ linearly extrapolated into the boundary and $v_{\phi}$ set using the stellar rotation rate $\Omega_{*}$, 
\begin{equation}
v_{\phi}=\Omega_{*}{r sin\theta}+\frac{\bf v_p \cdot B_p}{|{\bf B_p|}^2}B_{\phi},
\end{equation}
where the subscript ``p'' denotes the poloidal components ($r,\theta$) of a given vector. This condition enforces an effective rotation rate for the field lines which, in steady state ideal MHD, should be equal to the stellar rotation rate and conserved along field lines \citep{zanni2009mhd, reville2015effect}. This ensures the footpoints of the stellar magnetic field are correctly anchored into the surface of the star. The final boundary conditions are applied to the outer edges of the simulation, a simple outflow (zero derivative) is set at $60R_*$ allowing for the outward transfer of mass, momenta and magnetic field, along with an axisymmetric condition along the rotation axis ($\theta=0$ and $\pi$). Due to the supersonic flow properties at the outer boundary and its large radial extent compared with the location of the fast magnetosonic surface, any artefacts from the outer boundary cannot propagate upwind into the domain.  

The code is run, following the MHD equations above, until a steady state solution is found. The magnetic fields modify the wind dynamics compared to the spherically symmetric initial state, with regions of high magnetic pressure shutting off the radial outflow. In this way, the applied boundary conditions allow for closed and open regions of flow to form (e.g \citealp{washimi1993thermo}; \citealp{keppens2000stellar}), as observed within the solar wind. In some cases of strong magnetic field small reconnection events are seen, caused by the numerical diffusivity of our chosen numerical scheme. Reconnection events are also seen in Pantolmos \& Matt (in prep) and discussed within their Appendix. We adopt a similar method for deriving flow quantities in cases exhibiting periodic reconnection events. In such cases, once a quasi-steady state is established a temporal average of quantities such as the torque and mass loss are used. 

Inputs for the simulations are given as ratios of characteristic speeds which control key parameters such as the wind temperature ($c_s/v_{esc}$), field strength ($v_A/v_{esc}$) and rotation rate ($v_{rot}/v_{kep}$). Where $c_s=\sqrt{\gamma p/\rho}$ is the sound speed at the surface, $v_A=B_*/\sqrt{4\pi\rho}$ is the Alfv\'en speed at the north pole, $v_{rot}$ is the rotation speed at the equator, $v_{esc}=\sqrt{2GM_*/R_*}$ is the surface escape speed and $v_{kep}=\sqrt{GM_*/R_*}$ is the keplerian speed at the equator. In this way, all simulations represent a family of solutions for stars with a range of gravities. As this work focuses on the systematic addition of dipolar and quadrupolar geometries, we fix the rotation rate for all our simulations. \cite{matt2012magnetic} showed that the non-linear effects of rotation on their torque scaling can be neglected for slow rotators. They defined velocities as a fraction of the breakup speed,
\begin{equation}
f=\frac{v_{rot}}{v_{kep}}\bigg|_{r=R_*, \theta=\pi/2}=\frac{\Omega_*R_*^{3/2}}{(GM_*)^{1/2}}.
\end{equation}
The Alfv\'en radius remains independent of the stellar spin rate until $f\approx0.03$, after which the effects of fast rotation start to be important. For this study a solar rotation rate is chosen ($f=4.46\times10^{-3}$), which is well within the slow rotator regime. We set the temperature of the wind with $c_s/v_{esc}=0.25$, higher than $c_s/v_{esc}=0.222$ used previosuly in \cite{reville2015effect}. This choice of higher sound speed drives the wind to slightly higher terminal speeds, which are more consistent with observed solar wind speeds. Each geometry is studied with 10 different field strengths controlled by the input parameter $v_A/v_{esc}$, which is defined here with the Alfv\'en speed on the stellar north pole (see following Section). Table \ref{Parameters} lists all our variations of $v_A/v_{esc}$ for each geometry.

   \begin{figure*}
    \includegraphics[width=\textwidth]{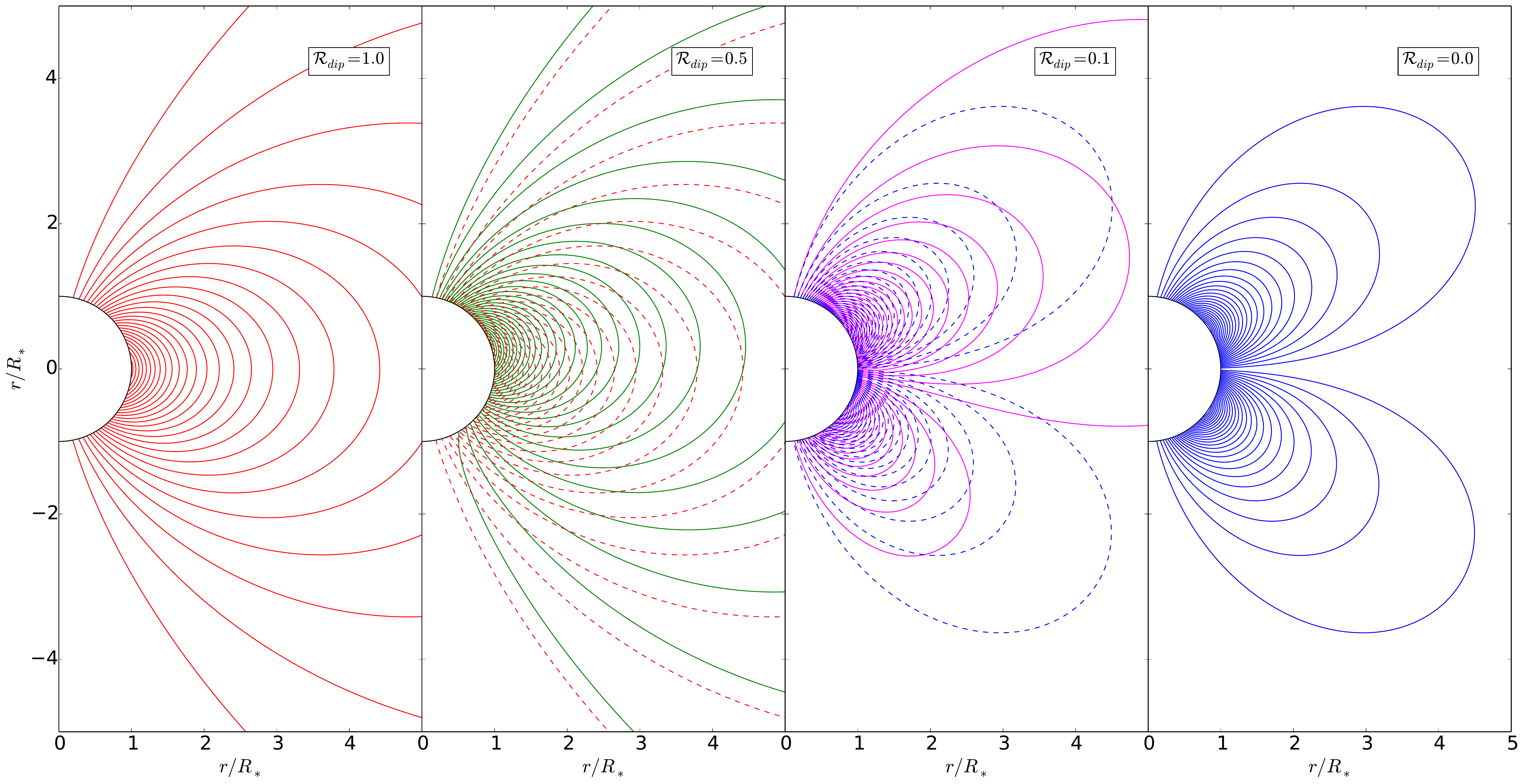}
     \caption{Initial magnetic configurations for a dipolar field, quadrupolar field and two mixed cases (red, green, magenta and blue for the dipole fractions of 100\%, 50\%, 10\% and purely quadrupole respectively). Mixed cases have the dominant pure field geometry over-plotted in dashed colour. The combined fields add in the northern hemisphere and subtract in southern hemisphere because they belong to opposite field symmetry families. With as much as half the field strength in the quadrupole, shown in green, the topology of the field is still dominated by the dipole field.}
     \label{RstudyField}
  \end{figure*}

Due to the use of characteristic speeds as simulation inputs, our results can be scaled to any stellar parameters. For example, using solar parameters, the wind is driven by a coronal temperature of $\approx$1.4MK and our parameter space covers a range of stellar magnetic field strengths from 0.9G to 87G over the pole. Changing these normalisations will modify this range.

\subsection{Magnetic Field Configuration}
  
Within this work, we consider magnetic field geometries that encompass a range of dipole and quadrupole combinations with different relative strengths. We represent the mixed fields using the ratio, $\mathcal{R}_{dip}$, of dipolar field to the total combined field strength. 

In this study the magnetic fields of the dipole and quadrupole are described in the formalism of \cite{gregory2010magnetic} using polar field strengths,
\begin{eqnarray}
B_{r,dip}(r,\theta) &=& B_{*}^{l=1} \left(\frac{R_*}{r}\right)^{3} \cos\theta, \\
B_{\theta,dip}(r,\theta) &=& \frac{1}{2} B_{*}^{l=1} \left(\frac{R_*}{r}\right)^{3} \sin\theta, \\
B_{r,quad}(r,\theta) &=& \frac{1}{2} B_{*}^{l=2} \left(\frac{R_*}{r}\right)^{4} (3\cos^2\theta -1) ,\\
B_{\theta,quad}(r,\theta) &=& B_{*}^{l=2} \left(\frac{R_*}{r}\right)^{4} \cos\theta\sin\theta.
\end{eqnarray}
The total field, comprised of the sum of the two geometries, 
\begin{equation}
{\bf B}(r,\theta) = {\bf B}_{dip}(r,\theta) + {\bf B}_{quad}(r,\theta),
\end{equation}
where the total polar field $B_*=B_*^{l=1}+B_*^{l=2}$, is controlled by the $\mathcal{R}_{dip}$ parameter,
\begin{equation}
\mathcal{R}_{dip}=\frac{B_{r,dip}}{B_{r,dip}+B_{r,quad}}\bigg|_{r=R_*,\theta=0}=\frac{B_*^{l=1}}{B_*}.
\label{R_dip}
\end{equation}
This work considers aligned magnetic moments such that $\mathcal{R}_{dip}$ ranges from 1 to 0, corresponding to all the field strength in the dipolar or quadrupolar mode respectively. As with $v_A/v_{esc}$, $\mathcal{R}_{dip}$ is calculated at the north pole. This sets the relative strengths of the dipole and quadrupole fields,
\begin{equation}
B_{*}^{l=1}=\mathcal{R}_{dip}B_{*}, \qquad B_{*}^{l=2}=(1-\mathcal{R}_{dip})B_{*},
\end{equation}

Alternative parametrisations are commonly used in the analysis of ZDI observations and dynamo modelling. These communities use the surface averaged field strengths, $\langle |B| \rangle$, or the ratio of magnetic energy density ($E_m\propto B^2$) stored within each of the dipole and quadrupole field modes at the stellar surface. During the solar magnetic cycle, values of $B^2_{quad}/B^2_{dip}$ can range from $\approx10-100$ at solar maximum to $\approx10^{-2}$ at solar minimum \citep{derosa2012solar}. A transformation from our parameter to the ratio of energies is simply given by:
\begin{equation}
\frac{B^2_{quad}}{B^2_{dip}}=\frac{2}{3}\frac{(1-\mathcal{R}_{dip})^2}{\mathcal{R}_{dip}^2},
\end{equation}
where the numerical pre-factor accounts for the integration of magnetic energy in each mode over the stellar surface.

   \begin{figure*}
    \includegraphics[width=\textwidth]{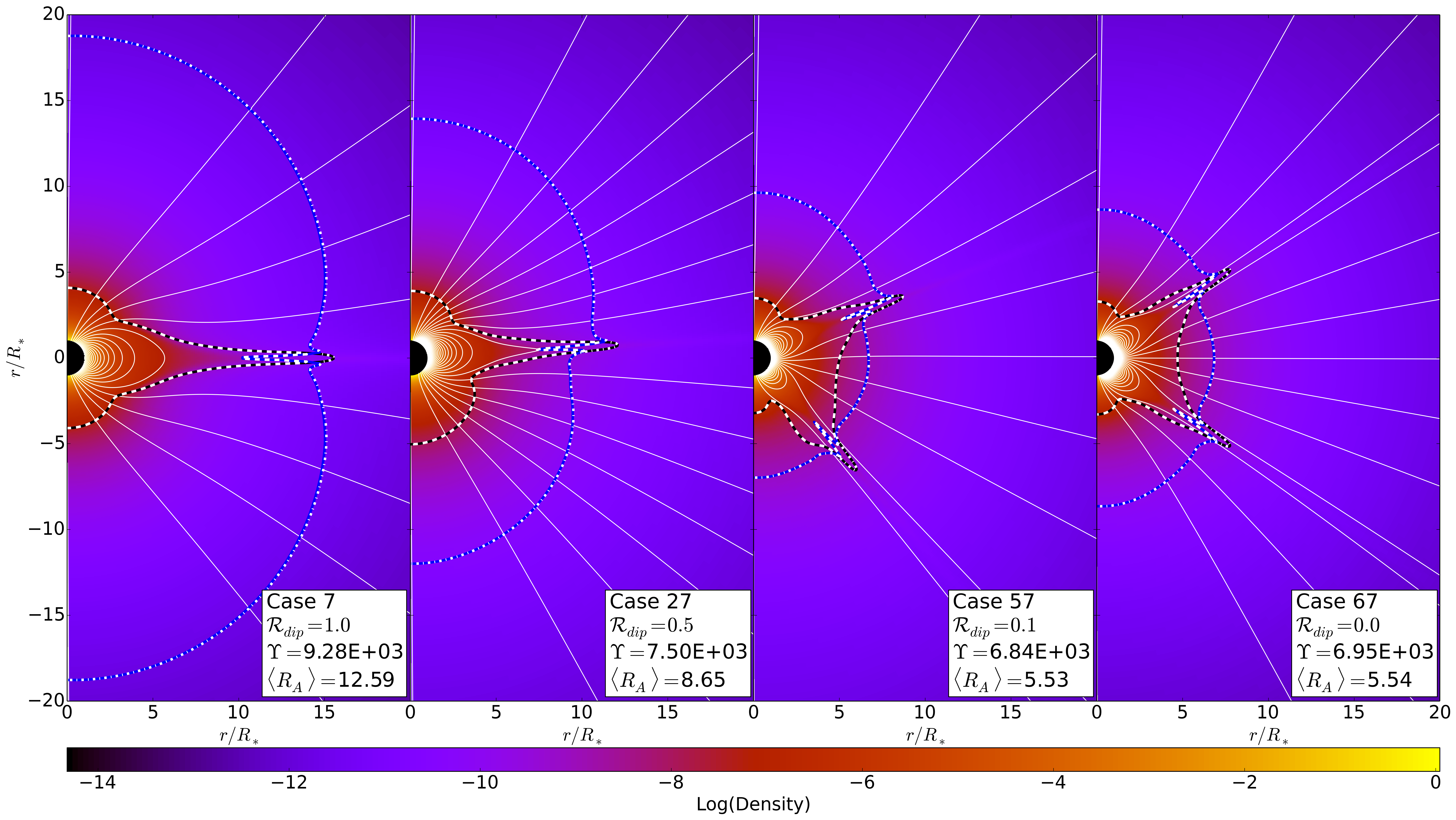}
     \caption{Logarithm of density normalised by the surface value for dipolar, quadrupolar and mixed magnetic fields for cases 7, 27, 57, 67 (see Table \ref{Parameters}). The winds are initialised using the same initial polytropic parker wind solution with $\gamma=1.05$ and $c_s/v_{esc}=0.25$. Stellar rotation rate and magnetic field strength are set with $f=4.46\times10^{-3}$ and $v_A/v_{esc}=3.0$. The Alfv\'en and sonic Mach surfaces are shown in blue and black respectively, in addition the fast and slow magnetosonic surfaces are indicated with dot-dash and dashed white lines. A transition from one to two streamers is seen with increasing quadrupolar field (decreasing $\mathcal{R}_{dip}$), and the two combined field cases exhibit the top bottom asymmetry from the field addition and subtraction.}
     \label{RstudyExample}
  \end{figure*}

Initial field configurations are displayed in Figure \ref{RstudyField}. The pure dipolar and quadrupolar cases are shown in comparison to two mixed cases ($\mathcal{R}_{dip}=0.5, 0.1$). These combined geometry fields add in one hemisphere and subtract in the other. This effect is due to the different symmetry families each geometry belongs to, with the dipole's polarity reversing over the equator unlike the equatorially symmetric quadrupole. Continuing the use of ``primary'' and ``secondary'' families as in \cite{mcfadden1991reversals} and \cite{derosa2012solar}, we refer to the dipole as primary and quadrupole as secondary. The fields are chosen such that they align in polarity in the northern hemisphere. This choice has no impact on the derived torque or mass loss rate due to the symmetry of the quadrupole about the equator. Either aligned or anti-aligned, these fields will always create one additive hemisphere and one subtracting; swapping their relative orientations simply switches the respective hemispheres. This is in contrast to combining dipole \& octupole fields, where the aligned and anti-aligned cases cause subtraction at the equator or poles respectively (\citealp{gregory2016multipolar}; Finley \& Matt. in prep). 

Figure \ref{RstudyField} indicates that even with equal quadrupole and dipole polar field strengths, $\mathcal{R}_{dip}=0.5$, the overall dipole topology will remain. In this case the magnetic energy density in the dipolar mode is 1.5 times greater than the quadrupolar mode and with the more rapid radial decay of the quadrupolar field, this explains the overall dipolar topology. A higher fraction of quadrupole is required to produce a noticeable deviation from this configuration, which is shown at $\mathcal{R}_{dip}=0.1$. More than half of the parameter space that we explore lies in the range where the energy density of the quadrupole mode is greater than that of the dipole ($B^2_{quad}/B^2_{dip}>1.0$). For this study both the pure dipolar and quadrupolar fields are used as controls (both of which were studied in detail within \cite{reville2015effect}), and 5 mixed cases parametrised by $\mathcal{R}_{dip}$ values ($\mathcal{R}_{dip}$ = 0.8, 0.5, 0.3, 0.2, 0.1). We include $\mathcal{R}_{dip}=0.8$ to demonstrate the dominance of the dipole at higher values. Each $\mathcal{R}_{dip}$ value is given a unique identifying colour which is maintained in all figures throughout this paper. Table \ref{Parameters} contains a complete list of parameters for all cases, which are numbered by increasing $v_A/v_{esc}$ and quadrupole fraction. 

\section{Simulation results}
\subsection{Morphology of the Field and Wind Outflow}

   \begin{sidewaysfigure*}
   \vspace*{-9cm} \includegraphics[width=\textwidth]{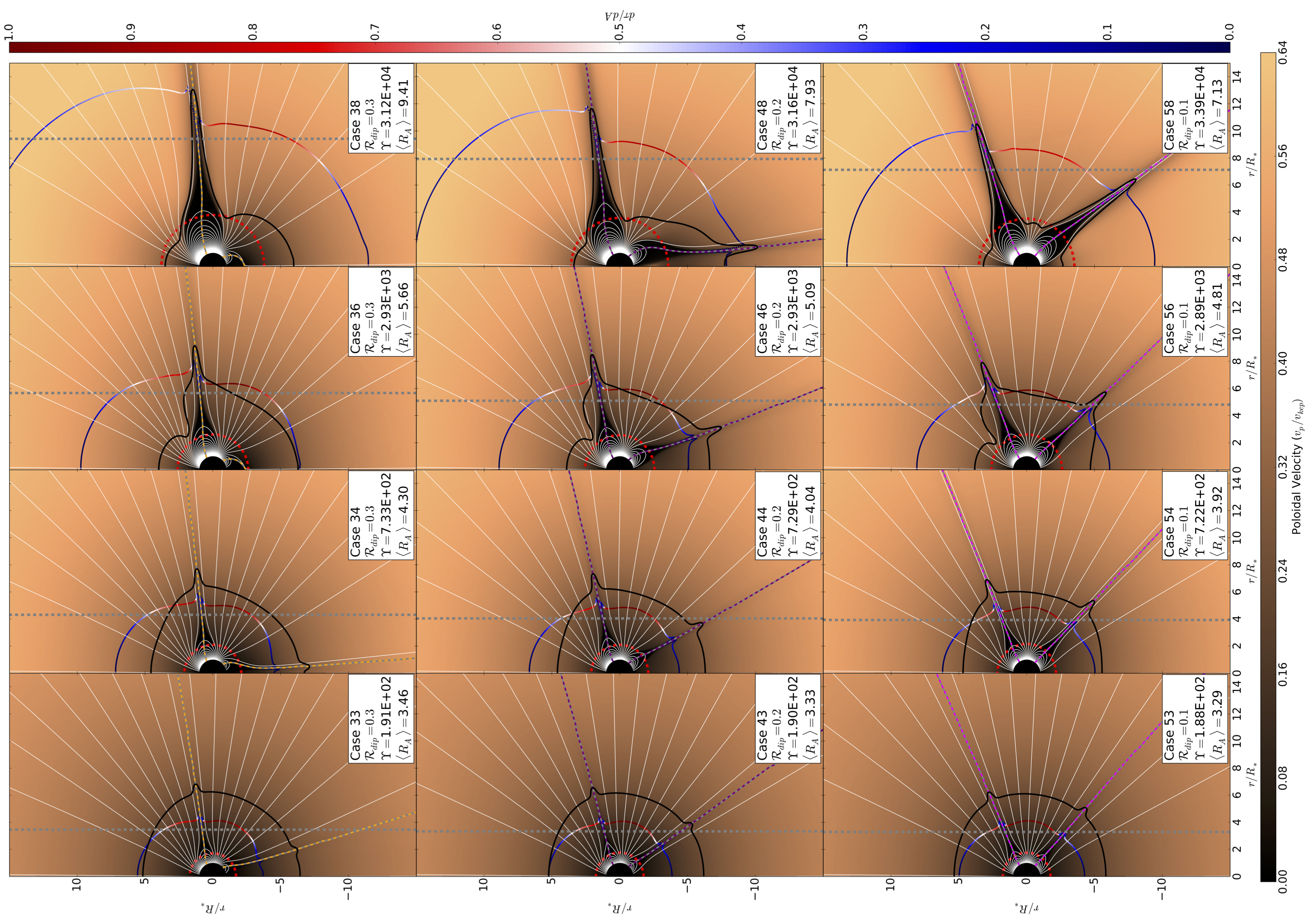}
     \caption{Simulation results for the lowest $\mathcal{R}_{dip}$ values 0.3, 0.2 and 0.1 (top, middle and bottom respectively), coloured by poloidal wind speed, with field lines in white. The current sheets are indicated by dashed lines, whose colour corresponds to their $\mathcal{R}_{dip}$ value in future figures. The streamer configuration is modified by both changes to the field strength and mixing ratio. Increased field strength or $\mathcal{R}_{dip}$ value tends to revolve the southern hemisphere streamer towards the south pole. The Alfv\'en surfaces have been coloured to show the flux of angular momentum normal to the surface [Units normalised by: $8\times10^{-6}\rho_*v_{kep}R_*$]. The average Alfv\'en radius, $\langle R_A \rangle$, from equation (\ref{R_A}) is shown in dashed grey. The sonic surface and opening radius are shown in solid black and dashed red respectively. The morphology and properties of the lower field cases are nearly indistinguishable, with only slight differences to the streamer locations. The reduction in torque with increasing quadrupolar fraction can be visually seen by moving down the grid. The most dipolar field sits in the top right panel and the most quadrupolar in the bottom left, these models are chosen to emphasis the transition in field dominance. }
     \label{streamers}
  \end{sidewaysfigure*}
  
Figure \ref{RstudyField} shows the topological changes in field structure from the addition of dipole and quadrupole fields. It is evident in these initial magnetic field configurations that the global magnetic field becomes asymmetric about the equator for mixed cases, as does the magnetic boundary condition which is maintained fixed at the stellar surface. It is not immediately clear how this will impact the torque scaling from \cite{reville2015effect}, who studied only single geometries.

Results for these field configurations using our PLUTO simulations are displayed in Figure \ref{RstudyExample}. The dipole and quadrupole cases are shown in conjunction with the mixed field cases, $\mathcal{R}_{dip}=0.5, 0.1$. The Figure displays for a comparable value of polar magnetic field strength, the different sizes of Alfv\'en surface that are produced. The mixed magnetic geometries modify the size and morphology of the Alfv\'en and sonic surfaces. Due to the slow rotation, the fast and slow magnetosonic surfaces are co-located with the sonic and Alfv\'en surfaces (the fast magnetosonic surface being always the larger of the two surfaces). 

The field geometry is found to imprint itself onto the stellar wind velocity with regions of closed magnetic field confining the flow creating areas of co-rotating plasma, referred to as deadzones \citep{mestel1968magnetic}. Steady state wind solutions typically have regions of open field where a faster wind and most of the torque is contained, along with these deadzone(s) around which a slower wind is produced. Similarly to the solar wind, slower wind can be found on the open field lines near the boundary of closed field \citep{feldman2005sources, riley2006comparison, fisk1998slow}. Observations of the Sun reveal the fast wind component emerging from deep within coronal holes, typically over the poles, and the slow wind component originating from the boundary between coronal holes and close field regions. Due to the polytropic wind used here, we do not capture the different heating and acceleration mechanisms required to create a true fast and slow solar-like wind \citep[as seen with the Ulysses spacecraft e.g.][]{mccomas2000solar, ebert2009bulk}. Our models produce an overall wind speed consistent with slow solar wind component, which we assume to represent the average global flow. More complex wind driving and coronal heating physics are required to recover a multi-speed wind, as observed from the Sun \citep{cranmer2007self, pinto2016flux}.

Figure \ref{streamers} displays a grid of simulations with a range of magnetic field strengths and $\mathcal{R}_{dip}=0.3, 0.2, 0.1$ values ($B^2_{quad}/B^2_{dip}$ ranges from 3.6 to 54; values consistent with the solar cycle maximum), where the mixing of the fields plays a clear role in the changing dynamics of the flow. Regions of closed magnetic field cause significant changes to the morphology of the wind. A single deadzone is established on the equator by the dipole geometry whereas the quadrupole creates two over mid latitudes. Mixed cases have intermediate states between the pure regimes. Within our simulations the deadzones are accompanied by streamers which form above closed field regions and drive slower speed wind than from the open field regions. The dynamics of these streamers, their location and size are an interesting result of the changing topology of the flow. 

The dashed coloured lines within Figure \ref{streamers} show where the field polarity reverses using $B_r=0$, which traces the location of the streamers. The motion of the streamers through the grid of simulations is then observed. With increasing quadrupole field, the single dipolar streamer moves into the northern hemisphere and with continued quadrupole addition a second streamer appears from the southern pole and travels towards the northern hemisphere until the quadrupolar streamers are recovered both sitting at mid latitudes. This motion can also be seen for fixed $\mathcal{R}_{dip}$ cases as the magnetic field strength is decreased. For a given $\mathcal{R}_{dip}$ value the current sheets sweep towards the southern hemisphere with increased polar field strength, in some cases (36 and 38) moving onto the axis of rotation. This is the opposite behaviour to decreasing the $\mathcal{R}_{dip}$ value, i.e. the streamer configuration is seen to take a more dipolar morphology as the field strength is increased. Additionally within Figure \ref{streamers}, for low field strengths each $\mathcal{R}_{dip}$ produces a comparable Alfv\'en surface with very similar morphology, all dominated by the quadrupolar mode. 

\subsection{Global Flow Quantities}
Our simulations produce steady state solutions for the density, velocity and magnetic field structure. To compute the wind torque on the star we calculate $\Lambda$, a quantity related directly to the angular momentum flux ${\bf F_{AM}}=\Lambda\rho{\bf v}$ \citep{keppens2000stellar},
\begin{equation}
\Lambda(r,\theta)=rsin\theta\bigg(v_{\phi}-\frac{B_{\phi}}{\rho}\frac{|{\bf B_p}|^2}{{\bf v_p \cdot B_p}}\bigg).
\end{equation}
Within axisymmetric steady state ideal MHD, $\Lambda$ is conserved along any given field line. However we find variations from this along the open-closed field boundary due to numerical diffusion across the sharp transition in quantities found there. The spin-down torque, $\tau$, due to the transfer of angular momentum in the wind is then given by the area integral,
\begin{equation}
\tau=\int_A\Lambda\rho{\bf v} \cdot d{\bf A},
\end{equation}
where $A$ is the area of any surface enclosing the star. For illustrative purposes, Figure \ref{streamers} shows the Alfv\'en surface coloured by angular momentum flux (thick multi-coloured line), which is seen to be strongly focused around the equatorial region. The angular momentum flux is calculated normal to the Alfv\'en surface,
\begin{equation}
\frac{d\tau}{dA}=\Lambda\rho{\bf v}\cdot {\bf \hat{A}}={\bf F_{AM}}\cdot {\bf \hat{A}},
\end{equation}
where ${\bf \hat{A}}$ is the normal unit vector to the Alfv\'en surface. The mass loss rate from our wind solutions is calculated similarly to the torque,
\begin{equation}
\dot{M}=\int_A\rho{\bf v} \cdot d{\bf A}.
\end{equation}
Both expressions for the mass loss and torque are evaluated using spherical shells of area $A$ which are outside the closed field regions. This allows for the calculation of an average Alfv\'en radius (which is cylindrical from the rotation axis) in terms of the torque, mass flux and rotation rate,
\begin{equation}
\langle R_A\rangle=\sqrt{\frac{\tau}{\dot{M}\Omega_*}}.
\label{R_A}
\end{equation}

   \begin{figure*}
    \includegraphics[width=\textwidth]{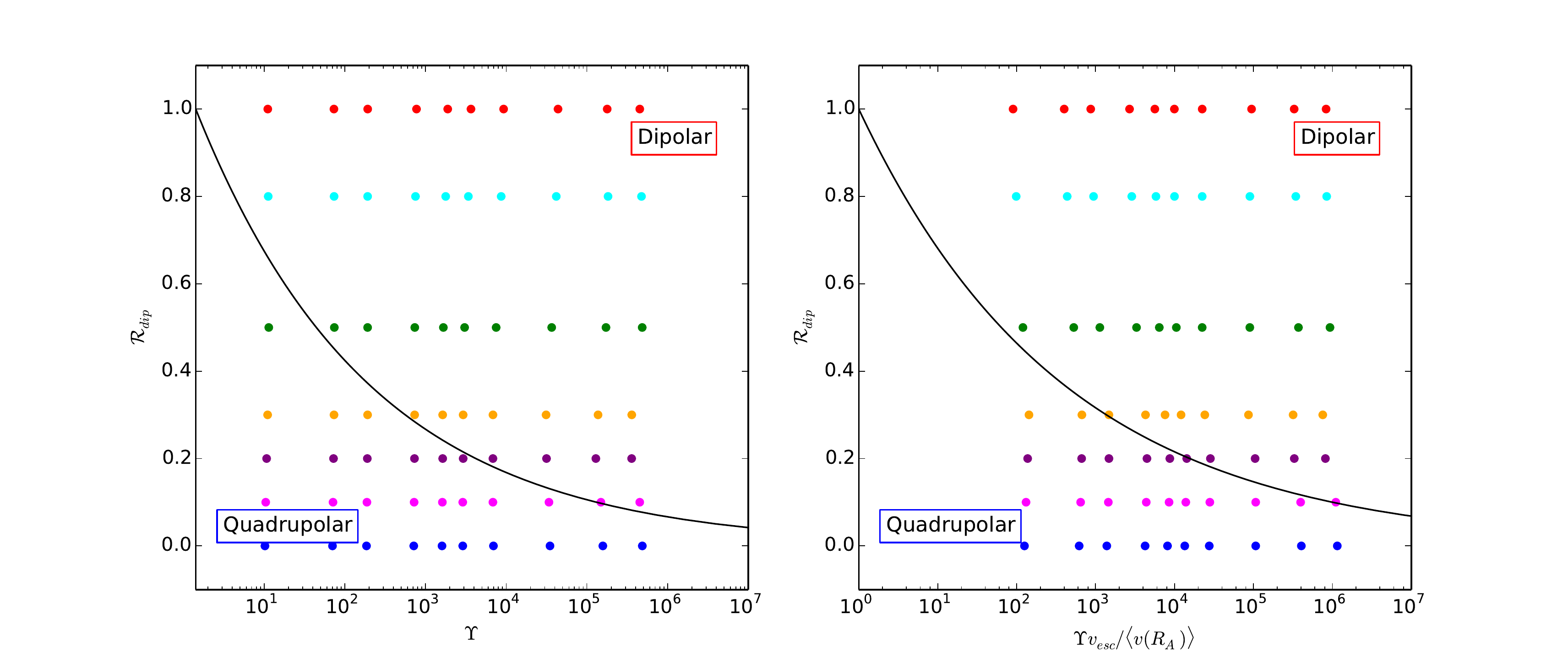}
     \caption{Parameter space explored in terms of $\Upsilon$, $\Upsilon v_{esc}/\langle v(R_A)\rangle$ and $\mathcal{R}_{dip}$. Five mixed geometries are explored along with pure cases of both dipole and quadrupole geometries. Colours for each $\mathcal{R}_{dip}$ value are used throughout this work. The black line indicates $\Upsilon_{crit}$, equation (\ref{UP_CRIT}). The formula for predicting the torque exhibits a quadrupolar scaling for $\Upsilon$ and $\mathcal{R}_{dip}$ values below the line, and dipolar above (See Section 3.4).}
     \label{PS}
  \end{figure*}
  
\begin{table*}
\caption{Input Parameters and Results from the 70 Simulations}
\label{Parameters}
\center
\setlength{\tabcolsep}{1pt}
    \begin{tabular}{cccccccc|cccccccc}
        \hline\hline
Case	&	$\mathcal{R}_{dip}$	&	$v_A/v_{esc}$	&	$\langle R_A\rangle/R_*$	&	$\Upsilon$	&	$R_o/R_*$	&	$\Upsilon_{open}$	&	$\langle v(R_A)\rangle/v_{esc} $	&	Case	&	$\mathcal{R}_{dip}$	&	$v_A/v_{esc}$	&	$\langle R_A\rangle/R_*$	&	$\Upsilon$	&	$R_o/R_*$	&	$\Upsilon_{open}$	&	$\langle v(R_A)\rangle/v_{esc} $	\\	\hline
1	&	1	&	0.1	&	3.06	&	11.1	&	1.31	&	294	&	0.123	&	36	&	0.3	&	2	&	5.66	&	2930	&	2.61	&	2040	&	0.242	\\	
2	&	1	&	0.3	&	4.19	&	73.2	&	1.88	&	819	&	0.183	&	37	&	0.3	&	3	&	6.76	&	6850	&	3.01	&	3460	&	0.283	\\	
3	&	1	&	0.5	&	5.05	&	192	&	2.33	&	1450	&	0.221	&	38	&	0.3	&	6	&	9.41	&	31200	&	3.8	&	8840	&	0.360	\\	
4	&	1	&	1	&	6.88	&	773	&	2.95	&	3550	&	0.287	&	39	&	0.3	&	12	&	13	&	137000	&	5.05	&	21600	&	0.432	\\	
5	&	1	&	1.5	&	8.56	&	1880	&	3.41	&	6530	&	0.334	&	40	&	0.3	&	24	&	15.7	&	360000	&	6.18	&	37300	&	0.476	\\	
6	&	1	&	2	&	10	&	3660	&	3.8	&	9970	&	0.367	&	41	&	0.2	&	0.1	&	2.43	&	10.7	&	1.2	&	120	&	0.078	\\	
7	&	1	&	3	&	12.6	&	9280	&	4.54	&	18100	&	0.414	&	42	&	0.2	&	0.3	&	2.96	&	72.4	&	1.54	&	245	&	0.109	\\	
8	&	1	&	6	&	18.2	&	43900	&	6.07	&	47000	&	0.463	&	43	&	0.2	&	0.5	&	3.33	&	190	&	1.76	&	368	&	0.129	\\	
9	&	1	&	12	&	25.1	&	178000	&	8	&	109000	&	0.544	&	44	&	0.2	&	1	&	4.04	&	729	&	2.1	&	701	&	0.163	\\	
10	&	1	&	24	&	29.6	&	452000	&	9.75	&	180000	&	0.543	&	45	&	0.2	&	1.5	&	4.61	&	1630	&	2.39	&	1070	&	0.187	\\	
11	&	0.8	&	0.1	&	2.51	&	11.2	&	1.2	&	245	&	0.114	&	46	&	0.2	&	2	&	5.09	&	2930	&	2.56	&	1480	&	0.205	\\	
12	&	0.8	&	0.3	&	3.89	&	73.5	&	1.76	&	651	&	0.168	&	47	&	0.2	&	3	&	5.92	&	6840	&	2.9	&	2390	&	0.240	\\	
13	&	0.8	&	0.5	&	4.64	&	192	&	2.1	&	1120	&	0.203	&	48	&	0.2	&	6	&	7.93	&	31600	&	3.58	&	5890	&	0.301	\\	
14	&	0.8	&	1	&	6.19	&	751	&	2.73	&	2620	&	0.261	&	49	&	0.2	&	12	&	10.4	&	129000	&	4.54	&	13500	&	0.392	\\	
15	&	0.8	&	1.5	&	7.6	&	1780	&	3.12	&	4690	&	0.305	&	50	&	0.2	&	24	&	12.6	&	359000	&	5.56	&	24500	&	0.439	\\	
16	&	0.8	&	2	&	8.88	&	3390	&	3.46	&	7210	&	0.339	&	51	&	0.1	&	0.1	&	2.44	&	10.5	&	1.2	&	121	&	0.079	\\	
17	&	0.8	&	3	&	11.1	&	8660	&	4.14	&	13100	&	0.386	&	52	&	0.1	&	0.3	&	2.95	&	71.3	&	1.54	&	243	&	0.110	\\	
18	&	0.8	&	6	&	16.3	&	41700	&	5.67	&	35000	&	0.463	&	53	&	0.1	&	0.5	&	3.29	&	188	&	1.76	&	358	&	0.129	\\	
19	&	0.8	&	12	&	22.9	&	183000	&	7.77	&	84500	&	0.531	&	54	&	0.1	&	1	&	3.92	&	722	&	2.16	&	652	&	0.164	\\	
20	&	0.8	&	24	&	27.9	&	475000	&	9.07	&	147000	&	0.560	&	55	&	0.1	&	1.5	&	4.41	&	1620	&	2.44	&	964	&	0.190	\\	
21	&	0.5	&	0.1	&	2.63	&	11.4	&	1.14	&	168	&	0.095	&	56	&	0.1	&	2	&	4.81	&	2890	&	2.61	&	1290	&	0.208	\\	
22	&	0.5	&	0.3	&	3.38	&	74.1	&	1.54	&	407	&	0.140	&	57	&	0.1	&	3	&	5.53	&	6840	&	2.9	&	2050	&	0.244	\\	
23	&	0.5	&	0.5	&	3.94	&	191	&	1.82	&	674	&	0.169	&	58	&	0.1	&	6	&	7.13	&	33900	&	3.52	&	4850	&	0.317	\\	
24	&	0.5	&	1	&	5.11	&	736	&	2.33	&	1500	&	0.223	&	59	&	0.1	&	12	&	8.96	&	149000	&	4.31	&	10300	&	0.376	\\	
25	&	0.5	&	1.5	&	6.11	&	1660	&	2.67	&	2510	&	0.259	&	60	&	0.1	&	24	&	10.5	&	452000	&	5.16	&	17700	&	0.408	\\	
26	&	0.5	&	2	&	7.03	&	3050	&	2.95	&	3740	&	0.289	&	61	&	0	&	0.1	&	2.47	&	10.2	&	1.2	&	127	&	0.081	\\	
27	&	0.5	&	3	&	8.65	&	7500	&	3.46	&	6670	&	0.334	&	62	&	0	&	0.3	&	2.98	&	70.3	&	1.59	&	256	&	0.113	\\	
28	&	0.5	&	6	&	12.6	&	36600	&	4.6	&	17900	&	0.407	&	63	&	0	&	0.5	&	3.33	&	185	&	1.82	&	377	&	0.134	\\	
29	&	0.5	&	12	&	18.3	&	172000	&	6.3	&	46000	&	0.464	&	64	&	0	&	1	&	3.96	&	715	&	2.22	&	682	&	0.168	\\	
30	&	0.5	&	24	&	23	&	485000	&	7.49	&	83300	&	0.519	&	65	&	0	&	1.5	&	4.44	&	1600	&	2.5	&	1010	&	0.196	\\	
31	&	0.3	&	0.1	&	2.46	&	11	&	1.14	&	124	&	0.077	&	66	&	0	&	2	&	4.83	&	2890	&	2.67	&	1350	&	0.214	\\	
32	&	0.3	&	0.3	&	3.04	&	73.4	&	1.48	&	268	&	0.109	&	67	&	0	&	3	&	5.54	&	6950	&	2.95	&	2150	&	0.252	\\	
33	&	0.3	&	0.5	&	3.46	&	191	&	1.71	&	420	&	0.130	&	68	&	0	&	6	&	6.98	&	34900	&	3.63	&	4910	&	0.326	\\	
34	&	0.3	&	1	&	4.3	&	733	&	2.1	&	870	&	0.171	&	69	&	0	&	12	&	8.46	&	158000	&	4.43	&	9970	&	0.390	\\	
35	&	0.3	&	1.5	&	5.03	&	1630	&	2.39	&	1420	&	0.215	&	70	&	0	&	24	&	9.65	&	584000	&	5.16	&	16400	&	0.421	\\	
        \hline
        \vspace{0.05cm}
    \end{tabular}
\end{table*}

Throughout this work, $\langle R_A\rangle$ is used as a normalised torque which accounts for the mass loss rates which we do not control. Values of the average Alfv\'en radius are tabulated within Table \ref{Parameters}. $\langle R_A \rangle$ is shown in Figure \ref{streamers} using a grey vertical dashed line. For each case, the cylindrical Alfv\'en radius is offset inwards of the maximum Alfv\'en radius from the simulation, a geometrical effect as this corresponds to the average cylindrical $R_A$ and includes variations in flow quantities as well. Exploring Figure \ref{streamers}, the motion of the deadzones/current sheets have little impact on the overall torque. For example, no abrupt increase in the Alfv\'en radius is seen from case 34 to 36 (where the southern streamer is forced onto the rotation axis) compared to cases 44 and 46. The torque is instead governed by the magnetic field strength in the wind which controls the location of the Alfv\'en surface.

We parametrise the magnetic and mass loss properties using the ``wind magnetisation'' defined by,
\begin{equation}
\Upsilon=\frac{B^2_*R_*^2}{\dot{M}v_{esc}},
\end{equation}
where $B_*$ is the combined field strength at the pole. Previous studies that used this parameter defined it with the equatorial field strength (e.g. \citealp{matt2008accretion}; \citealp{matt2012magnetic}; \citealp{reville2015effect}; Pantolmos \& Matt. in prep). We use polar values unlike previous authors due to the additive property of the radial field at the pole, for aligned axisymmetric fields. Note that selecting one value of the field on the surface will not always produce a value which describes the field as a whole. The polar strength works for these aligned fields, but will easily break down for un-aligned fields and anti-aligned axisymmetric odd $l$ fields, thus it suits the present study, but a move away from this parameter in future is warranted. 

During analysis, the wind magnetisation, $\Upsilon$, is treated as an independent parameter that determines the Alfv\'en radius $\langle R_A\rangle$ and thus the torque, $\tau$. We increase $\Upsilon$ by setting a larger $v_A/v_{esc}$, creating a stronger global magnetic field. Table \ref{Parameters} displays all the input values of $\mathcal{R}_{dip}$ and $v_A/v_{esc}$ as well as the resulting global outflow properties from our steady state solutions, which are used to formulate the torque scaling relations within this study. Figure \ref{PS} displays all 70 simulations in $\Upsilon-\mathcal{R}_{dip}$ space. Cases are colour-coded here by their $\mathcal{R}_{dip}$ value, a convention which is continued throughout this work.

\subsection{Single Mode Torque Scalings}

\begin{table*}
\caption{Best Fit Parameters to equations (\ref{UP_OLD}) and (\ref{UP_VA})}
\label{fitValues}
\center
    \begin{tabular}{ccc|ccc}
        \hline\hline
        Topology($l)$ & $K_s$   & $m_s$ &$K_l$   &$m_l$& $m_{l,th}(l)$   \\ \hline
        Dipole ($1$)    &  $1.49\pm0.03$  & $0.231\pm0.003$      &$0.92\pm0.04$ &$0.258\pm0.005$    &0.250           \\ 
        Quadrupole ($2$)    &  $1.72\pm0.03$  & $0.132\pm0.003$     &$1.11\pm0.04$&$0.156\pm0.004$&0.167                 \\   
        \hline
    \end{tabular}
\end{table*}

   \begin{figure*}
    \includegraphics[width=\textwidth]{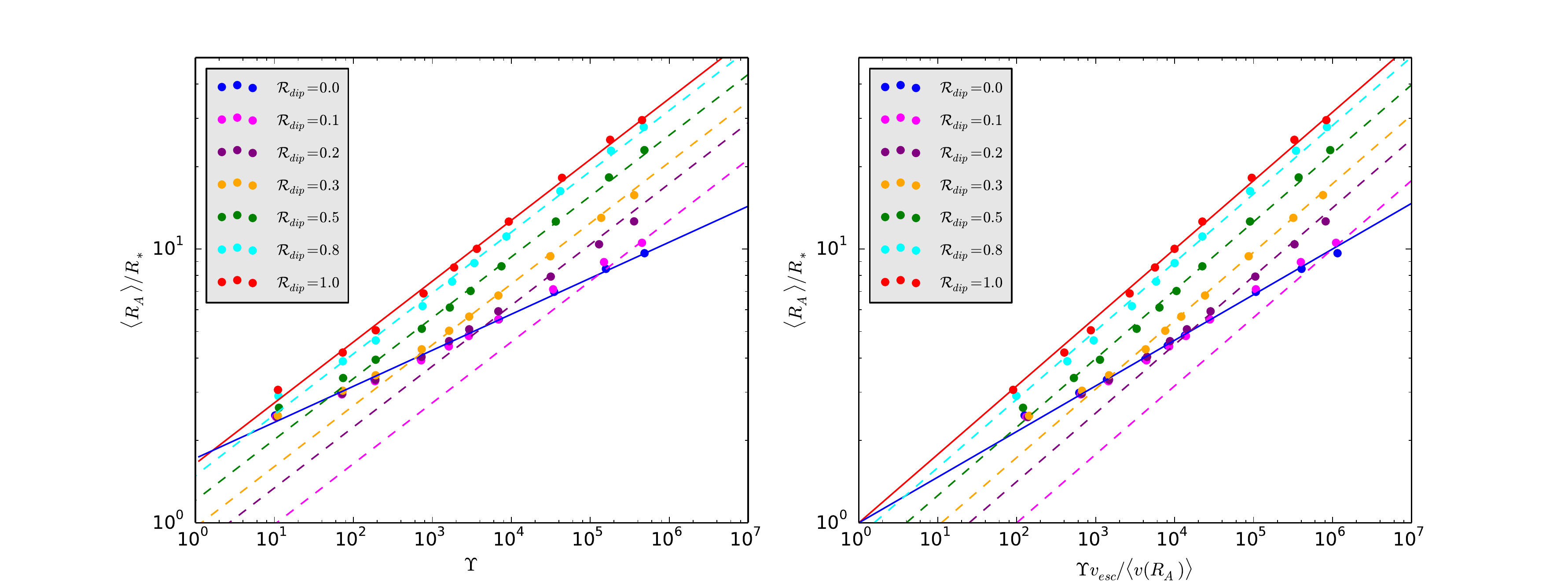}
     \caption{Average Alfv\'en radius vs wind magnetisation for all cases. Simulations are marked with colour-coded circles indicating their $\mathcal{R}_{dip}$ value. Left: Solid lines show the fit of dipole (red) and quadrupole (blue) to equation (\ref{UP_OLD}). Dashed lines show the dipolar component fit, equation (\ref{UP_DIPOLE}). Right: Solid lines show the analytic solution of dipole (red) and quadrupole (blue) to equation (\ref{UP_VA}) with $K_l=1$. Dashed lines show the dipolar component fit from equation (\ref{UP_DIPOLE2}), dependent on only the value of the field order $l$, unlike in the $\Upsilon$ space. }
     \label{RstudyUpsilon}
  \end{figure*}

The efficiency of the magnetic braking mechanism is known to be dependent on the magnetic field geometry. This has been previsously shown for single mode geometries \citep[e.g.][]{reville2015effect, garraffo2016missing}. We first concider two pure gemetries, dipole and quadrupole, using the formulation from \cite{matt2008accretion},
\begin{equation}
\frac{\langle R_A\rangle}{R_*} = K_{s} \Upsilon^{m_{s}},
\label{UP_OLD}
\end{equation}
where $K_s$ and $m_s$ are fitting parameters for the pure dipole and quadrupole cases, using the surface field strength. Here we empirically fit $m_s$; the interpretation of $m_s$ is discussed in \cite{matt2008accretion}, \cite{reville2015effect} and Pantolmos \& Matt (in prep), where it is determined to be dependant on magnetic geometry and the wind acceleration profile. The Appendix contains further discussion of the wind acceleration profile and its impact on this power law relationship. 

The left panel of Figure \ref{RstudyUpsilon} shows the Alfv\'en radii vs the wind magnetisations for all cases (colour-coded with their $\mathcal{R}_{dip}$ value). Solid lines show scaling relations for dipolar (red) and quadrupolar (blue) geometries, as first shown in \cite{reville2015effect}. We calculate best fit values for $K_s$ and $m_s$ for the dipole and quadrupole, tabulated in Table \ref{fitValues}. Values here differ due to our hotter wind ($c_s/v_{esc}=0.25$ than their $c_s/v_{esc}=0.222$), using polar $B_*$, and we do not account for our low rotation rate. As previously shown, the dipole field is far more efficient at transferring angular momentum than the quadrupole. In this study we concider the effect of combined geometries, within Figure \ref{RstudyUpsilon} these cases lie between the dipole and quadrupole slopes, with no single power law of this form to describe them.

Pantolmos \& Matt (in prep) have shown the role of the velocity profile in the power law dependence of the torque. In our simulations, the acceleration of the flow from the base wind velocity to its terminal speed is primarily governed by the thermal pressure gradient, however magnetic topologies can all modify the radial velocity profile (as can changes in wind temperature, $\gamma$, and rapid rotation, not included in our study). Effects on the torque formulations due to these differences in acceleration can be removed via the multiplication of $\Upsilon$ with $v_{esc}/\langle v(R_A)\rangle$. In their work, the authors determine the theoretical power law dependence, $m_{l,th}=1/(2l+2)$, from one-dimensional analysis. In this formulation the slope of the power law is controlled only by the order of the magnetic geometry, $l$, which is $l=1$ and $l=2$ for the dipole and quadrupole respectively,
\begin{equation}
\frac{\langle R_A\rangle}{R_*} = K_l\bigg[\Upsilon\frac{ v_{esc}}{\langle v(R_A)\rangle}\bigg]^{m_l},
\label{UP_VA}
\end{equation}
where $K_l$ and $m_l$ are fit parameters to our wind solutions, tabulated in Table \ref{fitValues}. The value of $\langle v(R_A)\rangle$ is calculated as an average of the velocity at all points on the Alfv\'en surface in the meridional plane. \footnote{It could be argued that this should be weighted by the total area of the Alfv\'en surface, but for simplicity we calculate the un-weighted average. }
 
Equation (\ref{UP_VA}) is able to predict accurately the power law dependence for the two pure modes using the order of the spherical harmonic field, $l$. We show this in the right panel of Figure \ref{RstudyUpsilon}, where the Alfv\'en radii are plotted against the new parameter, $\Upsilon v_{esc}/\langle v(R_A)\rangle$. A similar qualitative behaviour is shown to the scaling with $\Upsilon$ in the left panel. Using the theoretical power law dependencies, the  dipolar (red) and quadrupolar (blue) slopes are plotted with $m_{l,th}=1/4$ and $m_{l,th}=1/6$ respectively. Using a single fit constant $K_l=1$ for both sloes within this figure shows good agreement with the simulation results. 

More accurate values of $K_l$ and $m_l$ are fit for each mode independently. These values produce a better fit and are compared with the theoretical values in Table \ref{fitValues}. The mixed simulations show a similar qualitative behaviour to the plot against $\Upsilon$. 

Obvious trends are seen within the mixed case scatter. A saturation to quadrupolar Alfv\'en radii values for lower $\Upsilon$ and $\mathcal{R}_{dip}$ values is observed, along with a power law trend with a dipolar gradient for higher $\Upsilon$ and $\mathcal{R}_{dip}$ values. This indicates that both geometries play a role in governing the lever arm, with the dipole dominating the braking process at higher wind magnetisations.

\subsection{Broken Power Law Scaling For Mixed Field Cases}

   \begin{figure}
    \includegraphics[width=0.5\textwidth]{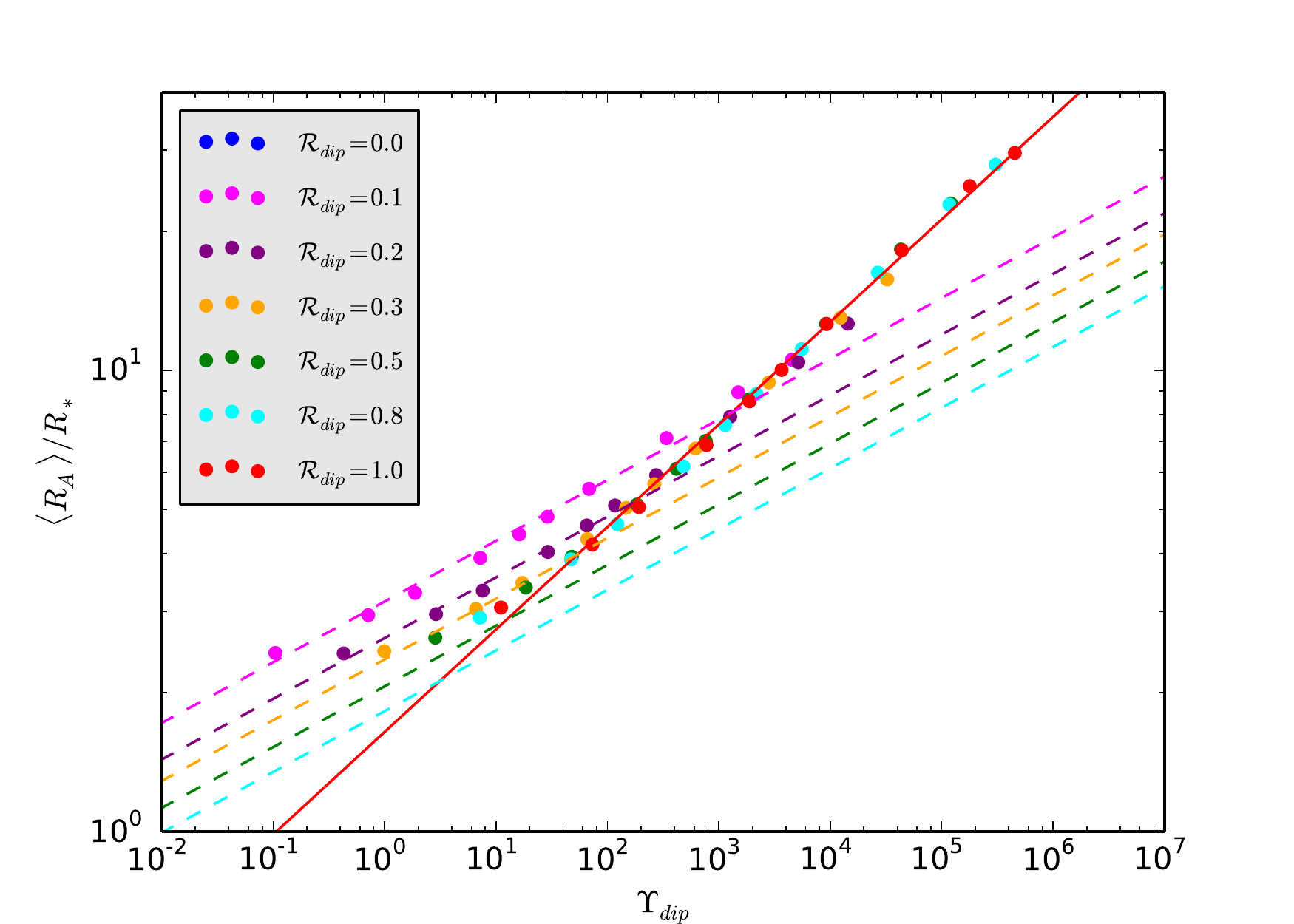}
     \caption{Average Alfv\'en radius vs the dipolar wind magnetisation. Considering only the dipolar field strength, we produce a single power law for the Alf\'ven radius, equation (\ref{UP_DIPOLE}). Our wind solutions are shown to agree well with dipole prediction in most cases. Disagreement at low $\Upsilon_{dip}$ and $\mathcal{R}_{dip}$ values are explained by the quadrupolar slopes, shown in coloured dashed lines.}
     \label{Upsilon_dip}
  \end{figure}

Observationally the field geometries of cool stars are, at large scales, dominated by the dipole mode with higher order $l$ modes playing smaller roles in shaping the global field. It is the global field which controls the spin-down torque in the magnetic braking process. Higher order modes (such as the quadrupole) decay radially much faster than the dipole and as such they have a reduced contribution to setting the Alfv\'en speed at distances larger than a few stellar radii. 

We calculate $\Upsilon_{dip}$, which only takes into account the dipole's field strength,
\begin{equation}
\Upsilon_{dip}=\bigg(\frac{B^{l=1}_*}{B_{*}}\bigg)^2\frac{B^2_{*}R_*^2}{\dot{M}v_{esc}}=\mathcal{R}_{dip}^2\Upsilon.
\end{equation}
Taking as a hypothesis that the field controlling the location of the Alfv\'en radius is the dipole component, a power law scaling using $\Upsilon_{dip}$ can be constructed in the same form as Matt \& Pudritz (2008),
\begin{equation}
\frac{\langle R_A\rangle}{R_*} = K_{s,dip}[\Upsilon_{dip}]^{m_{s,dip} } = K_{s,dip}[\mathcal{R}_{dip}^2\Upsilon]^{m_{s,dip} }.
\label{UP_DIPOLE}
\end{equation}
Substitution of the dipole component into equation (\ref{UP_VA}) similarly gives,
\begin{equation}
\frac{\langle R_A\rangle}{R_*} = K_{l,dip}\bigg[\mathcal{R}_{dip}^2\Upsilon_{}\frac{ v_{esc}}{\langle v(R_A)\rangle}\bigg]^{m_{l,dip}},
\label{UP_DIPOLE2}
\end{equation}
where $K_{s,dip}$, $m_{s,dip}$, $K_{l,dip}$, and $m_{l,dip}$ will be parameters fit to simulations.

A comparison of these approximations can be seen in Figure \ref{RstudyUpsilon}, where equations (\ref{UP_DIPOLE}) (left panel) and (\ref{UP_DIPOLE2}) (right panel) are plotted with dashed lines for all the $\mathcal{R}_{dip}$ values used in our simulations. Mixed cases which lie above the quadrupolar slope are shown to agree with the dashed-lines in both forms. Such cases are dominated by the dipole component of the field only, irrespective of the quadrupolar component. 

The role of the dipole is even more clear in Figure \ref{Upsilon_dip} where only the dipole component of $\Upsilon$ is plotted for each simulation. The solid red line in Figure \ref{Upsilon_dip}, given by equation (\ref{UP_DIPOLE}), shows agreement at a given $\mathcal{R}_{dip}$ with deviation from this caused by a regime change onto the quadrupolar slope (shown in dashed colour). 

The behaviour of our simulated winds, despite using a combination of field geometries, simply follow existing scaling relations with this modification. In general, the dipole ($\Upsilon_{dip}$) prediction shows good agreement with the simulated wind models, except in cases where the Alfv\'en surface is close-in to the star. In these cases, the quadrupole mode still has magnetic field strength able to control the location of the Alfv\'en surface. Interestingly, and in contrast to the dipole-dominated regime, the quadrupole dominated regime behaves as if all the field strength is within the quadrupolar mode. This is visible within Figure \ref{RstudyUpsilon} for low values of $\Upsilon$ and $\mathcal{R}_{dip}$.

The mixed field $\langle R_A \rangle$ scaling can be described as a broken power law, set by the maximum of either the dipole component or the pure quadrupolar relation. With the break in the power law given by $\Upsilon_{crit}$,
\begin{equation}
  \frac{\langle R_A \rangle}{R_*}=\left\{
  \begin{array}{@{}ll@{}}
    K_{s,dip}[\mathcal{R}_{dip}^2\Upsilon]^{m_{s,dip}}, & \text{if}\ \Upsilon>\Upsilon_{crit}(\mathcal{R}_{dip}), \\
    K_{s,quad}[\Upsilon]^{m_{s,quad}}, & \text{if}\ \Upsilon\leq\Upsilon_{crit}(\mathcal{R}_{dip})
  \end{array}\right.
  \label{BrokenPowerLaw}
\end{equation} 
where $\Upsilon_{crit}$ is the location of the intercept for the dipole component and pure quadrupole scalings,
\begin{equation}
\Upsilon_{crit}(\mathcal{R}_{dip})=\bigg[\frac{K_{s,dip}}{K_{s,quad}}\mathcal{R}_{dip}^{2m_{s,dip}} \bigg]^{\frac{1}{m_{s,quad}-m_{s,dip}}}.
\label{UP_CRIT}
\end{equation}
The solid lines in Figure \ref{PS} show the value of $\Upsilon_{crit}$, equation (\ref{UP_CRIT}), diving the two regimes. Specifically, the solutions above the solid black line behave as if only the dipole component ($\Upsilon_{dip}$) is governing the Alfv\'en radius.

Transitioning from regimes is not perfectly abrupt. Therefore producing an analytical solution for the mixed cases which includes this behaviour would increase the accuracy for stars near the regime change. E.g. we have formulated a slightly better fit, using a relationship based on the quadrature addition of different regions of field. However it provides no reduction to the error on this simpler form and is not easily generalised to higher topologies. For practical purposes, the scaling of equation (\ref{BrokenPowerLaw}) and (\ref{UP_CRIT}) predict accurately the simulation torque with increasing magnetic field strength for a variety of dipole fractions. We therefore present the simplest available solution, leaving the generalised form to be developed within future work. 

   \begin{figure*}
    \includegraphics[width=\textwidth]{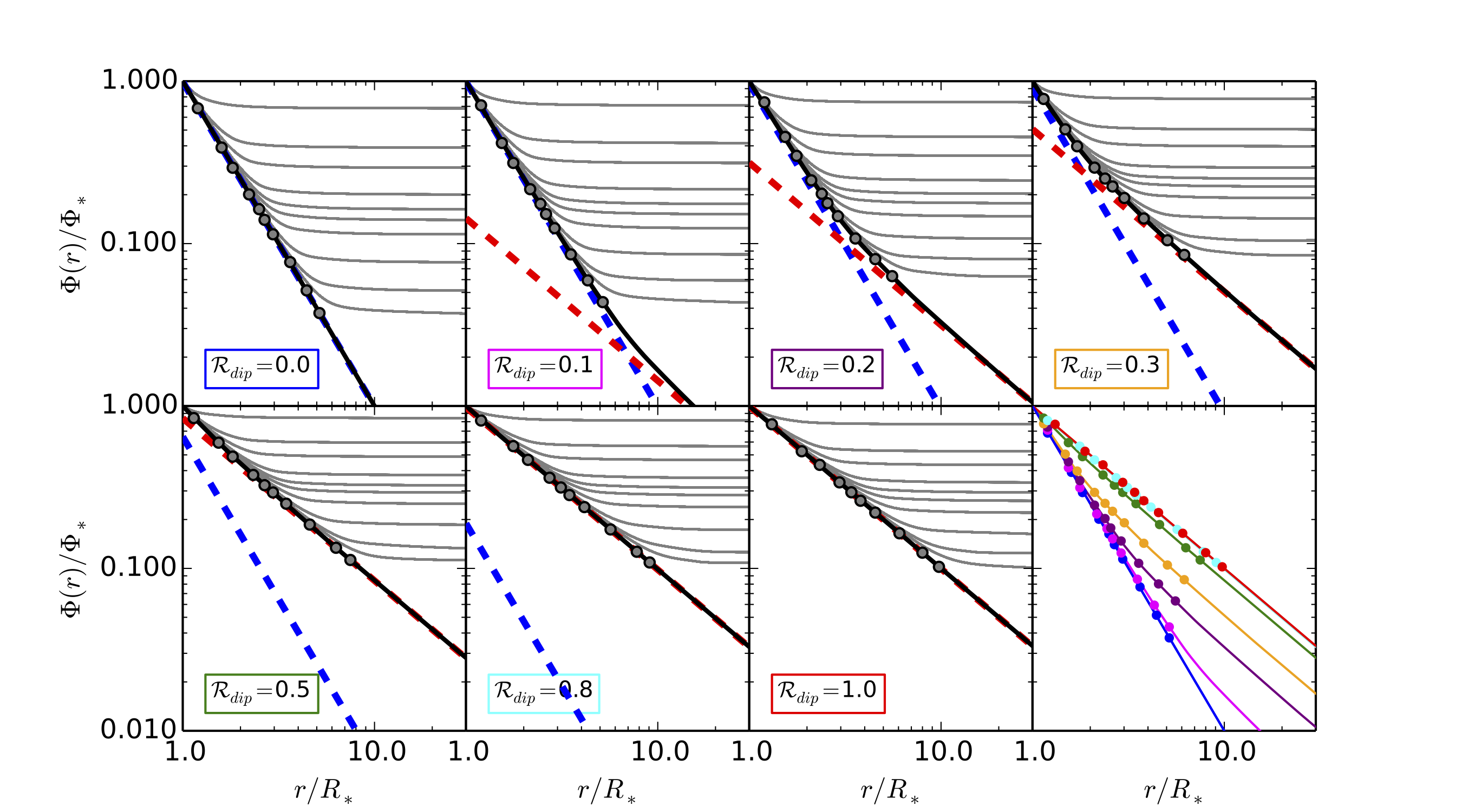}
     \caption{Magnetic flux vs Radial distance for all cases studied within this work, compared with analytical predictions. Solid grey lines show the 10 simulation fields for each field geometry. Solutions of equation (\ref{phi}) for the potential field magnetic flux are shown in black solid lines for each $\mathcal{R}_{dip}$ value.  In each case, the flux of dipole and quadrupole components using a potential field are plotted with dashed red and blue respectively, equations (\ref{quad_flux}) and (\ref{dip_flux}). Each simulation matches the potential field flux, until the wind pressures open the field to a constant flux. The open flux radii are displayed as grey circles. The lower right panel shows a comparison of each potential field flux decay along with the opening radii for each case (i.e. the solid black lines and grey circles from the other panels), colour-coded to the value of $\mathcal{R}_{dip}$.}
     \label{RstudyRaRo}
  \end{figure*}
  
\section{The impact of geometry on the magnetic flux in the wind} 
\subsection{Evolution of the Flux}
The magnetic flux in the wind is a useful diagnostic tool. The rate of the stellar flux decay with distance is controlled by the overall magnetic geometry. We calculate the magnetic flux as a function of radial distance by evaluating the integral of the magnetic field threading closed spherical shells, where we take the absolute value of the flux to avoid field polarity cancellations,
\begin{equation}
\Phi(r)=\oint_r|{\bf B} \cdot d{\bf A}|.
\label{phi}
\end{equation}
Considering the initial potential fields of the two pure modes this is simply a power law in field order $l$,
\begin{equation}
\Phi(r)_P=\Phi_*\bigg(\frac{R_*}{r}\bigg)^l,
\label{potential}
\end{equation}
where $l=1$ dipole and $l=2$ quadrupole, we denote the flux with ``$P$'' for the potential field. Figure \ref{RstudyRaRo} displays the flux decay of all values of $v_A/v_{esc}$ for each $\mathcal{R}_{dip}$ value, grey lines. The behaviour is qualitatively identical to that observed within previous works \citep[e.g.][]{schrijver2003asterospheric, johnstone2010modelling, vidotto2014m, reville2015effect}, where the field decays as the potential field does until the pressure of the wind forces the field into a purely radial configuration with a constant magnetic flux, referred to as the open flux. The power law dependence of equation (\ref{potential}) indicates for higher $l$ mode magnetic fields, the decay will be faster. We therefore expect the more quadrupolar dominated fields studied in this work to have less open flux. 

In the case of mixed geometries a simple power law is not available for the initial potential configurations, instead we evaluate the flux using equation (\ref{phi}), where $\bf B$ is the initial potential field for each mixed geometry. This allows us to calculate the radial evolution of the flux for a given $\mathcal{R}_{dip}$ which we compare to the simulated cases. Figure \ref{RstudyRaRo} shows the flux normalised by the surface flux versus radial distance from the star. For each $\mathcal{R}_{dip}$ value, the magnetic flux decay of the potential field (black solid line) is shown with the different strength $v_A/v_{esc}$ simulations (grey solid lines). A comparison of the flux decay for all potential magnetic geometries is available in the bottom right panel showing, as expected, the increasingly quadrupolar fields decaying faster.

In this study we control $v_A/v_{esc}$ which, for a given surface density, sets the polar magnetic field strength for our simulations. The stellar flux for different topologies and the same $B_*$ will differ and must be taken into account in order to describe the dipole and quadrupolar components (dashed red and blue) in Figure \ref{RstudyRaRo}. We plot the magnetic flux of the potential field quadrupole component alone in dotted blue for each $\mathcal{R}_{dip}$ value,
\begin{equation}
\Phi(r)_{P,quad}=(1-\mathcal{R}_{dip})\Phi_{*,quad}\bigg(\frac{R_*}{r}\bigg)^2,
\label{quad_flux}
\end{equation}
and similarly the potential field dipole component of the magnetic flux,
\begin{equation}
\Phi(r)_{P,dip}=\mathcal{R}_{dip}\Phi_{*,dip}\bigg(\frac{R_*}{r}\bigg),
\label{dip_flux}
\end{equation}
where in both equations the surface flux of a pure dipole/quadrupole ($\Phi_{*,dip}$, $\Phi_{*,quad}$) field is required to match our normalised flux representation. 

Due to the rapid decay of the quadrupolar mode, the flux at large radial distances for all simulations containing the dipole mode is described by the dipolar component. The quadrupole component decay sits below and parallel to the potential field prediction for small radii, becoming indistinguishable for the lowest $\mathcal{R}_{dip}$ values as the flux stored in the dipole is decreased. Importantly for small radii, simulations containing a quadrupolar component are dominated by the quadrupolar decay following a $l=2$ power law decay, which can be seen by shifting the blue dashed line upwards to intercept $\Phi/\Phi_*=1$ at the stellar surface. 

This result for the flux decay is reminiscent of the broken power law description for the Alfv\'en radius in Section 3.4. The field acts as a quadrupole using the total field for small radii and the dipole component only for large radii. There is a transition between these two regimes that is not described by either approximation. But is shown by the potential solution in solid black. 

\subsection{Topology Independent Open Flux Formulation}
  
The magnetic flux within the wind decays following the potential field solution closely until the magnetic field geometry is opened by the pressures of the stellar wind and the field lines are forced into a nearly radial configuration with constant flux, shown in Figure \ref{RstudyRaRo} for all simulations. The importance of this open flux is discussed by \cite{reville2015effect}. These authors showed a single power law dependence for the Alfv\'en radius, independent of magnetic geometry, when parametrised in terms of the open flux, $\Phi_{open}$,
\begin{equation}
\Upsilon_{open}=\frac{\Phi_{open}^2/R_*^2}{\dot{M}v_{esc}},
\end{equation}
which, ignoring the effects of rapid rotation, can be fit with,
\begin{equation}
\frac{\langle R_A\rangle}{R_*} = K_o[\Upsilon_{open}]^{m_o},
\label{UP_OPEN_OLD}
\end{equation}
where, $m_o$ and $K_o$ are fitting parameters for the open flux formulation. 

\begin{table*}
\caption{Open Flux Best Fit Parameters to equations (\ref{UP_OPEN_OLD}) and (\ref{UP_OPEN})}
\label{fitValues_open}
\center
    \begin{tabular}{ccccc}
        \hline\hline
        Topology($l)$ &  $K_o$& & $m_o$   \\ \hline
        Dipole ($1$)    &  $0.37\pm0.05$  && $0.360\pm0.006$             \\ 
        Quadrupole ($2$)                & $0.62\pm0.01$  & &$0.283\pm0.002$              \\   
        \hline
         &    $K_c$ & $K_{c,th}$& $m_c$ & $m_{c,th}$  \\ \hline
        Topology Independent & $0.08\pm0.03$&$ 0.0796$&$0.471\pm0.003$&0.500\\
        \hline
    \end{tabular}
\end{table*}

   \begin{figure*}
    \includegraphics[width=\textwidth]{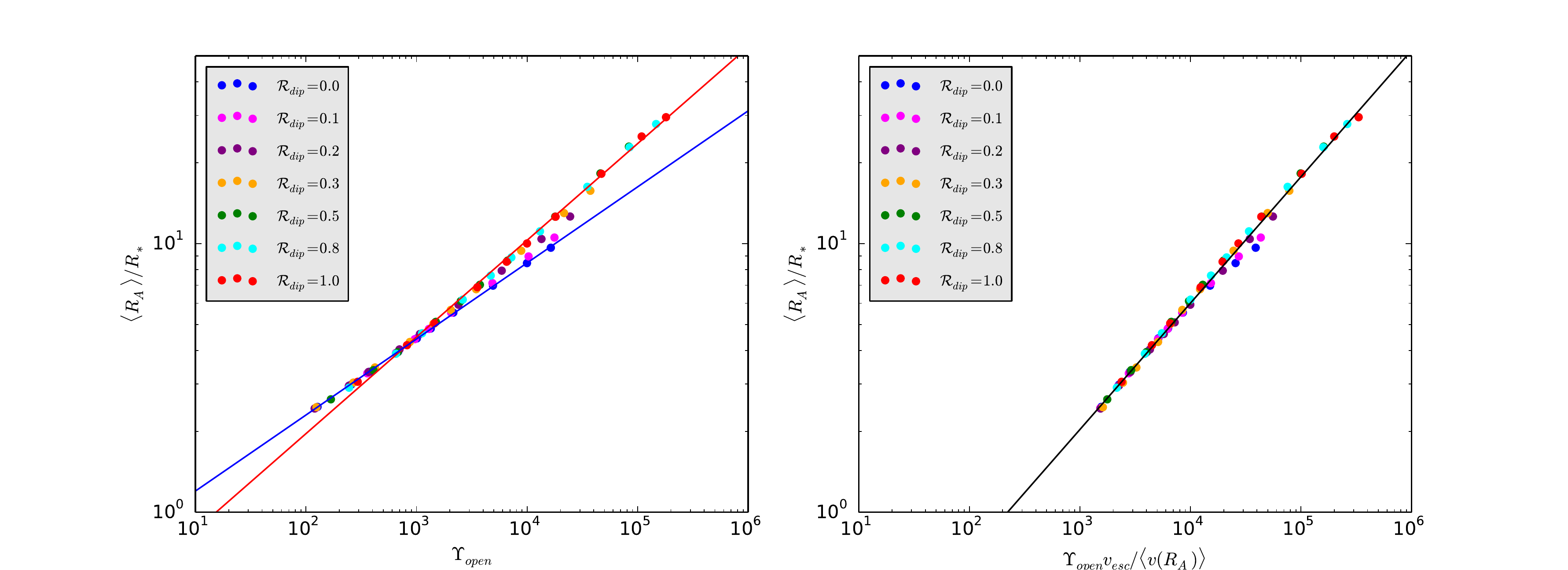}
     \caption{Left: Average Alfv\'en radius vs open flux magnetisation for all cases.  Fits to equation (\ref{UP_OPEN_OLD}) are shown for the dipole ($\mathcal{R}_{dip}=1$) and quadrupole ($\mathcal{R}_{dip}=0$)  fields. The geometry of the field is shown to influence the scaling relation, due to differences in the wind acceleration. Right: Average Alfv\'en radius vs open flux magnetisation accounting for the acceleration profile using work done by Pantolmos \& Matt (in prep). The fit of equation (\ref{UP_OPEN}) is shown to reduce the scatter for all simulations. A systematic discrepancy is still seen from the single power law with changing geometry.}
     \label{RstudyUpsilonOpen}
  \end{figure*}

Using the open flux parameter, Figure \ref{RstudyUpsilonOpen} shows a collapse towards a single power law dependence as in \cite{reville2015effect}. However our wind solutions show a systematic difference in power law dependence from dipole to quadrupole. On careful inspection of the result from Figure 6 of \cite{reville2015effect}, the same systematic trend between their topologies and the fit scaling is seen.  \footnote{A choice in our parameter space may have made this clearer to see in Figure \ref{RstudyUpsilonOpen}, due to the increased heating and therefore larger range of acceleration allowing the topology to impact the velocity profile.} We calculate best fits for each pure mode separately i.e. the dipole and quadrupole, tabulated in Table \ref{fitValues_open}.

Pantolmos \& Matt (in prep) find solutions for thermally driven winds with different coronal temperatures, from these they find the wind acceleration profiles of a given wind to very significantly alter the slope in $R_A$-$\Upsilon_{open}$ space. From this work our trend with geometry indicates that each geometry must have a slightly different wind acceleration profile. This is most likely due to difference in the super radial expansion of the flux tubes for each geometry, which is not taken into account with equation (\ref{UP_OPEN_OLD}). The field geometry is imprinted onto the wind as it accelerates out to the Alfv\'en surface. As such, this scaling relation is not entirely independent of topology. Further details on the wind acceleration profile within our study is available in the Appendix. Pantolmos (in prep) are able to include the effects of acceleration in their scaling through multiplication of $\Upsilon_{open}$ with $v_{esc}/\langle v(R_A)\rangle$. The expected semi-analytic solution from Pantolmos \& Matt (in prep) is given,
\begin{equation}
\frac{\langle R_A\rangle}{R_*} = K_c\bigg[\Upsilon_{open}\frac{ v_{esc}}{\langle v(R_A)\rangle}\bigg]^{m_c},
\label{UP_OPEN}
\end{equation}
where the fit parameters are derived from one-dimensional theory as constants, $K_{c,th}=1/4\pi$ and $m_{c,th}=1/2$.

We are able to reproduce this power law fit of $\Upsilon_{open}$ with the wind acceleration effects removed, on the right panel of Figure \ref{RstudyUpsilonOpen}. Including all simulations in the fit, we arrive at values of $K_c=1.01K_{c,th}\pm0.07$ and $m_c=0.942m_{c,th}\pm0.009 $ for the constants of proportionality and power law dependence. However a systematic difference is still seem from one $\mathcal{R}_{dip}$ value to another. More precise fits can be found for each geometry independently, but the systematic difference appearing in the right panel implies a modification to our semi-analytic formulations is required to describe the torque fully in terms of the open flux.

Here we show the scaling law from \cite{reville2015effect} is improved with the modification from Pantlomos (in prep). This formulation is able to describe the Alfv\'en radius scaling with changing open flux and mass loss. However with the open flux remaining an unknown from observations and difficult to predict, scaling laws that incorporate known parameters (such as those of equations (\ref{BrokenPowerLaw}) and (\ref{UP_CRIT})) are still needed for rotational evolution calculations. 

\subsection{The Relationship Between the Opening and Alfv\'en Radii}

   \begin{figure}
    \includegraphics[width=0.5\textwidth]{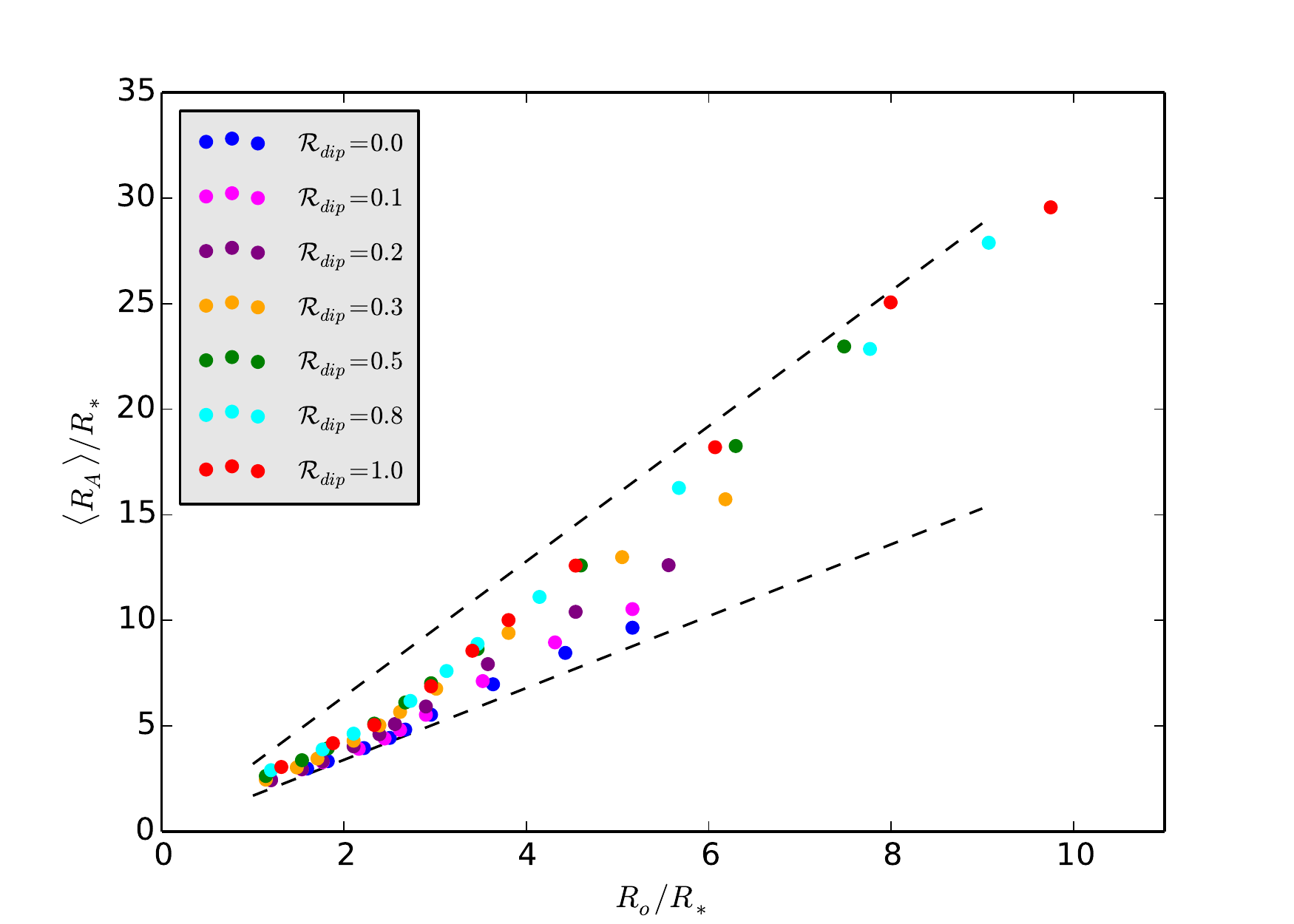}
     \caption{Alfv\'en radii vs opening radii for all simulated cases. Black dashed line represent $R_A/R_o=3.2$ and $1.7$.  Different geometries have a changing relationship between the torque lever arm and the opening radius of the field.}
     \label{RstudyRaRoAll}
  \end{figure}
  
The location of the field opening is an important distance. It is both critical for determining the torque and for comparison to potential field source surface (PFSS) models \citep[][]{altschuler1969magnetic}, which set the open flux with a tunable free parameter $R_{ss}$. The opening radius, $R_o$, we define is the radial distance at which the potential flux reaches the value of the open flux ($\Phi_P(R_o)=\Phi_{open}$). This definition is chosen because it relates to the 1D analysis employed to describe the power law dependences of our torque scaling relations. Specifically, a known value of $R_o$ allows for a precise calculation of the open flux (a priori from the potential field equations), which then gives the torque on the star within our simulations. The physical opening of the simulation field takes place at slightly larger radii than this with the field becoming non-potential due to its interaction with the wind (which explains why the closed field regions seen in Figure \ref{streamers} typically extend slightly beyond $R_o$). A similar smooth transition is produced with PFSS modelling.” 

$R_o$ is marked for each simulation in Figure \ref{RstudyRaRo} and again for comparative purposes in the bottom right panel. It is clear that smaller opening radii are found for lower $\mathcal{R}_{dip}$ cases. Due to their more rapidly decaying flux, they tend to have a smaller fraction of the stellar flux remaining in the open flux. From the radial decay of the magnetic field, the open flux and opening radii are observed to be dependent on the available stellar flux and topology. Pantolmos \& Matt (in prep) have recently shown these to also be dependent on the wind acceleration profile. This complex dependence makes it difficult to predict the open flux for a given system. 

Our simulations produce values for the average Alfv\'en radius, $\langle R_A\rangle$, and the opening radius, $R_o$, for the 7 different geometries studied. It is interesting to consider the relative size of these radii as they both characterise key dynamic properties for each stellar wind solution. For all cases shown in Figure \ref{streamers}, the opening radii are plotted in dashed red, allowing for the relative size to be compared with the cylindrical Alfv\'en radius, shown in dash grey. With increasing magnetic field strength ($\Upsilon$), both radii are seen to grow from case to case, however with increasing $\mathcal{R}_{dip}$, the cylindrical Alfv\'en radius generally grows faster than the opening radius. To quantify this, Figure \ref{RstudyRaRoAll} shows a plot of the Alfv\'en radii vs the opening radii for all cases. Linear trends of $R_A/R_o=3.2$ and $1.7$ are indicated with dashed lines.  For each $\mathcal{R}_{dip}$ value, the relationship between the Alfv\'en and opening radius ($\langle R_A\rangle/R_o$) is seen to systematically decrease with increasing higher order field component. In all cases, for small radii a shallower slope is observed which then steepens with increasing radial extent. 

The dependence of the Alfv\'en radius and opening radius on field geometry and magnetisation is a constraint on PFSS models, which are readily used with ZDI observations as a less computationally expensive alternative to MHD modelling \citep{jardine1999potential, jardine2002global, dunstone2008first, cohen2010coronal, johnstone2010modelling, rosen2015first, reville2015solar}. PFSS models are a useful tool, however require the source surface radius, $R_{ss}$, as an input. Authors often set a source surface and change the geometry and strength of the field freely \citep{fares2010searching, see2015energy, see2016studying}. We find however for a given $\mathcal{R}_{dip}$ value there exists a differing relation for the opening radius, as we define it here, to the Alfv\'en radius and magnetisation. These trends are observed to continue for higher $l$ mode fields (Finley \& Matt. in prep), with $\langle R_A\rangle/R_o$ decreasing overall with increased field complexity. As such, our results confirm that the opening radius should not remain fixed when changing geometries or increasing the wind magnetisation. We find the relationship of $\langle R_A\rangle/R_o$ to change in both cases. With fixed magnetisation, the opening radius should move towards the star for higher order fields to maintain a constant thermal driving. Maintaining the opening radius whilst increasing the field complexity infers that the wind has a reduced acceleration. Similarly with increased wind magnetisation the opening radius should move further from the star. The value of $R_o$ as we have defined it, is directly related to the source surface radius, and for a given magnetic geometry, the two should scale approximately together. For example, for a dipole field, comparing our definition of $R_o$ to the PFSS model shows that $R_{ss}$ equals an approximately constant value of 3/2 $R_o$. Thus conclusions made about the opening radii are constraints on future PFSS modelling. 

A method for predicting $R_o$ within our simulations remains unknown, however it is understood that $R_o$ is key to predicting the torque from our simulated winds. We do however find the ratio of $\langle R_A\rangle/R_o$ to be roughly constant for a given geometry, deviations from which may be numerical or suggest additional physics which we do not explore here.

\section{Conclusion}
We undertake a systematic study of the two simplest magnetic geometries, dipolar and quadrupolar, and for the first time their combinations with varying relative strengths. We parametrise the study using the ratio, $\mathcal{R}_{dip}$, of dipolar to total combined field strength, which is shown to be a key variable in our new torque formulation. 

We have shown that a large proportion of the magnetic field energy needs to be in the quadrupole for any significant morphology changes to be seen in the wind. All cases above 50\% dipole field show a single streamer and are dominated by dipolar behaviour. Even in cases of small $\mathcal{R}_{dip}$ we observe the dipole field to be the key parameter controlling the morphology of the flow, with the quadrupolar field rapidly decaying away for most cases leaving the dipole component behind. For smaller field strengths the Alfv\'en radii appears close to the star, where the quadrupolar field is still dominant, and thus a quadrupolar morphology is established. Increasing the fraction of quadrupolar field strength allows this behaviour to continue for larger Alfv\'en radii.

The morphology of the wind can be concidered in the context of star-planet or disk interactions. Our findings suggest that the connectivity, polarity and strength of the field within the orbital plane depend in a simple way on the relative combination of dipole and quadrupole fields. Different combinations of these two field modes change the location of the current sheet(s) and the relative orientation of the stellar wind magnetic field with respect to any planetary or disk magnetic field. Asymmetries such as these can modify the poynting flux exchange for close-in planets \citep{strugarek2014diversity} or the strength of magnetospheric driving and geomagnetic storms on Earth-like exoplanets. \cite{cohen2014magnetospheric} use observed magnetic fields to simulate the stellar wind environment surrounding the planet hosting star EV Lac. They calculate the magnetospheric joule heating on the exoplanets orbiting the M dwarf, finding significant changes to atmospheric properties such as thickness and temperature. Additionally, transient phenomena in the Solar wind such as coronal mass ejections are shown to deflect towards streamer belts \citep{kay2013forecasting}. This has been applied to mass ejections around M dwarfs stars \citep{kay2014implications}, and could similarly be applied here using knowledge of the streamer locations from our model grid. 

If the host star magnetic field can be observed and decomposed into constituent field modes, containing dominant dipole and quadrupole components, a qualitative assessment of the stellar wind environment can be made. We find the addition of these primary and secondary fields to create an asymmetry which may shift potentially habitable exoplanets in and out of volatile wind streams. Observed planet hosting stars such as $\tau$ Bootis have already been shown to have global magnetic fields which are dominated by combinations of these low order field geometries \citep{donati2008magnetic}. With further investigation it is possible to qualitatively approximate the conditions for planets in orbit of such stars. For dipole and quadrupole dominated host stars with a given magnetic field strength our grid of models provide an estimate of the location of the streamers and open field regions.

Within this work we build on the scaling relations from, \cite{matt2012magnetic}, \cite{reville2015effect} and Pantolmos \& Matt (in prep). We confirm existing scaling laws and explore a new mixed field parameter space with similar methods. From our wind solutions we fit the variables, $K_{s,dip}$, $m_{s,dip}$, $K_{s,quad}$ and $m_{s,quad}$ (see Table \ref{qandms}), which describe the torque scaling for the pure dipole and quadrupole modes. From the 50 mixed case simulations, we produce an approximate scaling relation which takes the form of a broken power law, as a single power law fit is not available for the mixed geometries cases in $\Upsilon$ space. 

For low $\Upsilon$ and low dipole fraction, the Alfv\'en radius behaves like a pure quadrupole,
\begin{equation}
\tau =  K_{s,quad} \dot{M} \Omega_*R_*^2[ \Upsilon]^{2m_{s,quad}} ,
\end{equation}
\begin{equation}
 = K_{s,quad} \dot{M}^{1-2m_{s,quad}} \Omega_*R_*^{2+4m_{s,quad}}\bigg[\frac{(B_{*})^2}{v_{esc}}\bigg]^{2m_{s,quad}}.
\end{equation}
At higher $\Upsilon$ and dipole fractions, the torque is only dependent on the dipolar component of the field,
\begin{equation}
\tau =  K_{s,dip} \dot{M} \Omega_*R_*^2[ \Upsilon_{dip}]^{2m_{s,dip}} ,
\end{equation}
\begin{equation}
 = K_{s,dip} \dot{M}^{1-2m_{s,dip}} \Omega_*R_*^{2+4m_{s,dip}}\bigg[\frac{(B^{l=1}_{*})^2}{v_{esc}}\bigg]^{2m_{s,dip}}.
\end{equation}
The later formulation is used when the Alfv\'en radius of a given dipole \& quadrupole mixed field is greater than the pure quadrupole case for the same $\Upsilon$, i.e. the maximum of our new formula or the pure quadrupole. We define $\Upsilon_{crit}$ to separate the two regimes (see Figure \ref{PS}).

The importance of the relative radial decay of both modes and the location of the opening and Alfv\'en radii appear to play a key role, and deserve further follow up investigation. This work analytically fits the decay of the magnetic flux, but a parametric relationship for the field opening remains uncertain. The relation of the relative sizes of the Alfv\'en and opening radii are found to be dependent on geometry, which can be used to inform potential field source surface modelling, where by the source surface must be specified when changing the field geometry. 

Paper II includes the addition of octupolar field geometries, another primary symmetry family which introduces an additional complication in the relative orientation of the octupole to the dipole. It is shown however, that the mixing of any two axisymmetric geometries will follow a similar behaviour, especially if each belongs to different symmetry families (Finley \& Matt. in prep). The lowest order mode largely dominates the dynamics of the torque until the Alfv\'en radii and opening radii are sufficiently close to the star for the higher order modes to impact the field strength.



\acknowledgments
Thanks for helpful discussions and technical advice from Georgios Pantolmos \& Matt, Victor See, Victor R\'eville, Sasha Brun and Claudio Zanni.
This project has received funding from the European Research Council (ERC) under the European Union’s Horizon 2020 research and innovation programme (grant agreement No 682393).
We thank Andrea Mignone and others for the development and maintenance of the PLUTO code.
Figures within this work are produced using the python package matplotlib \citep{hunter2007matplotlib}.





\appendix
\section{Wind Acceleration}

\begin{table*}[h!]
\caption{Predicting $m_s$ and $m_o$ using $q=0.8\pm0.1$}
\label{qandms}
\center
    \begin{tabular}{c|cc|cc}
        \hline\hline
        Topology($l)$ & $m_s$ & $m_{s,th}(l,q)$ & $m_o$ & $m_{o,th}(q)$ \\ \hline
        Dipole ($1$)    &  $0.231\pm0.003$   &$0.21\pm0.01$   & $0.360\pm0.006$   & $0.36\pm0.02$     \\ 
        Quadrupole ($2$)    &  $0.132\pm0.003$   &$0.15\pm0.01$  & $0.283\pm0.002$   & $0.36\pm0.02$  \\   
        \hline
    \end{tabular}
\end{table*}

   \begin{figure}[h!]
   \center
    \includegraphics[width=0.6\textwidth]{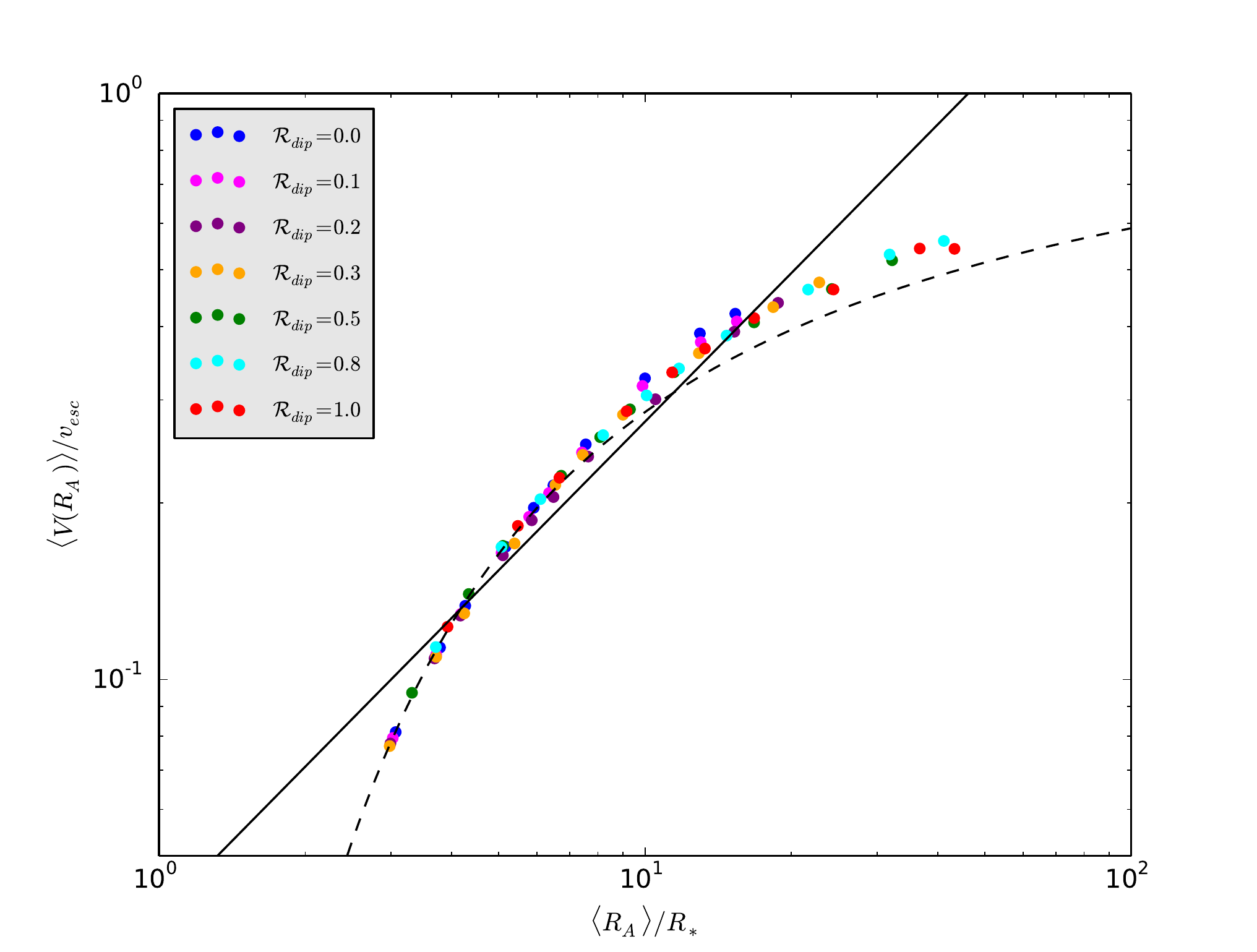}
     \caption{Scatter of the average Alfv\'en speed at the Alfv\'en surface as a function of the average Alfv\'en radius. The dashed line shows a hydrodynamic parker wind with $c_s/v_{esc}=0.25$, the solid line shows a fit to all our simulation data. Variation is seen between the dipolar and quadrupolar data towards the extreme values of the Alfv\'en radius. The combined average wind acceleration profile in black gives a $q=0.84$. The winds in our simulations are set with a higher coronal temperature than \cite{reville2015effect}, thus show a larger acceleration (they produce $q\approx0.7$).}
     \label{VA_RA}
  \end{figure}

The creation of a semi-analytic formulation for the Alfv\'en radius for a variety of stellar parameters has been the goal of many studies proceeding this (e.g. \citealp{matt2008accretion}; \citealp{matt2012magnetic}; \citealp{reville2015effect}; Pantolmos \& Matt. in prep). Using a one-dimensional approximation based on work by \cite{kawaler1988angular}, previous studies have aimed to predict the power law dependence, $m$, of the torque formulations used within this work. 

Using the one-dimensional framework, the field strength is assumed to decay as a power law $B(r)=B_*(R_*/r)^{l+2}$, which in this study is only valid for the pure cases. Pantolmos \& Matt (in prep) show the effect of wind acceleration can be removed from the torque scaling relations through the multiplication of $\Upsilon$ and $\Upsilon_{open}$ with $v_{esc}/\langle v(R_A)\rangle$. The power law dependences then becomes,
\begin{equation}
m_{l,th}=1/(2l+2),
\end{equation} 
and similarly,
\begin{equation}
m_{c,th}=1/2.
\end{equation}
The modified dependent parameter, $\Upsilon v_{esc}/\langle v(R_A)\rangle$, is used throughout this work (see Figures \ref{RstudyUpsilon} and \ref{RstudyUpsilonOpen}), and the analytic predictions for the power law slopes are shown to have good agreement with our simulations. This dependent variable however, requires additional information about the wind speed at the Alfv\'en surface which is often unavailable.

Typically, rotation evolution models use the available stellar surface parameters e.g. $\Upsilon$. Therfore knowledge of the flow speed at the Alfv\'en radius, $v(R_A)$, is required for the semi-analytic formulations. $v(R_A)$ is shown by Pantolmos \& Matt. (in prep) and \cite{reville2015effect} to share a similar profile to a one-dimensional thermal wind, $v(r)$. Figure \ref{VA_RA} displays the average Alfv\'en speed vs the Alfv\'en radius for all 70 simulations (coloured points). The parker wind solution \citep{parker1965dynamical} used in the initial condition is displayed for comparison (dashed line). Nearly all simulations follow the hydrodynamic solution, with a behaviour mostly independent of $\mathcal{R}_{dip}$. Towards higher values of the Alfv\'en radius, a noticeable separation starts to develop between geometries. This range is accessed less by the higher $l$ order geometries as the range of Alfv\'en radii is much smaller than that for the pure dipole mode.

In order to include the effects of wind acceleration in the simplified 1D analysis to explain the simulation scalings between $R_A$ and $\Upsilon$, \cite{reville2015effect} introduced a parametrisation for the acceleration of the wind to the Alfv\'en radius with a power law dependence in radial distance using $q$,
\begin{equation}
v(R_A)/v_{esc}=(R_A/R_*)^q .
\label{q}
\end{equation}
A single power law with $q=0.84$ is fit to the simulation data, which is chosen for simplicity within the 1D formalism. The use of this $q$ parameter is approximate if $v(R_A)$ is a power law in $R_A$, which we show over the parameter space has a significant deviation. Using the semi-analytic theory, \cite{reville2015effect} then derived the power law dependence for the $\Upsilon$ scaling (Equation (\ref{UP_OLD})),
\begin{equation}
m_{s,th}=1/(2l+2+q) ,
\end{equation}
which includes geometric and wind acceleration parameters in the form of $l$ and $q$ respectively. Using this result, $m_{s,th}$ is computed for both the dipole ($l=1$) and quadrupole ($l=2$) geometries in Table \ref{qandms}, and compared to the simulation results with good agreement.

Pantolmos \& Matt. (in prep) explain the power-law dependence, so long as $R_o/R_A$ remains constant and the wind acceleration profile is known. \cite{reiners2012radius}, \cite{reville2015effect} and Pantolmos \& Matt. (in prep) all analytically describe the power law dependence of the open flux formulation (Equation (\ref{UP_OPEN_OLD})) using the power law dependence $q$,
\begin{equation}
m_{o,th}=1/(2+q) . 
\end{equation}
The result is independent of geometry, $l$. As before the $q$ parameter approximates the wind driving as a power law in radius, which is fit with a single power law for both geometries such that $m_{o,th}$ should be the same for both the dipole and quadrupole. This prediction is tabulated in Table \ref{qandms}, however the simulation slopes are shown to no longer agree with the result. It is suggested that the open flux slope is much more sensitive to the wind acceleration than the $\Upsilon$ formulation, therefore slight changes in flow acceleration modify the result. Slightly different slopes can be fit for the dipole and quadrupole cases which can recover the different $m_o$ values, however this is seemingly just a symptom of the power law approximation breaking down. 

We conclude that the approximate power law of equation (\ref{q}) give a reasonable adjustment to the torque prediction for known wind velocity profiles, despite the badness of fit to the simulations points. Even though the power-law approximation to the wind velocity profile (equation \ref{q}) is not a precise fit to the data in Figure \ref{VA_RA}, the value of $q$ does provide a way to approximately include the contribution of the wind acceleration to the fit power-law exponents $m_o$ and $m_s$. A more precise formulation could be derived based on a Parker-like wind profile without the use of a power law, however the torque scaling with $\Upsilon$ is relative insensitive to the chosen approximate velocity profile. \\





\bibliographystyle{yahapj}
\bibliography{Paper1}




\end{document}